\documentclass{elsart}

\usepackage{graphicx}

\usepackage{amssymb}
\usepackage{ifpdf}
\usepackage{bm}
\usepackage{amsmath,epsfig}
\usepackage{tabularx}

\newcommand{\abs}[1]{\left\vert#1\right\vert}

\begin{document}

\begin{frontmatter}



\title{Correlation, hierarchies, and networks in financial markets}


\author[label1]{Michele Tumminello}
\author[label1,label2]{Fabrizio Lillo}
\author[label1]{Rosario N. Mantegna}
\address[label1]{Dipartimento di Fisica e Tecnologie Relative, Universit\`a di Palermo, Viale delle Scienze, I-90128 Palermo, Italy}
\address[label2]{Santa Fe Institute, 1399 Hyde Park Road, Santa Fe, NM 87501, U.S.A.}

\begin{abstract}
We discuss some methods to quantitatively investigate the properties of correlation matrices. Correlation matrices play an important role in portfolio optimization and in several other quantitative descriptions of asset price dynamics in financial markets. Specifically, we discuss how to define and obtain hierarchical trees, correlation based trees and networks from a correlation matrix. The hierarchical clustering and other procedures performed on the correlation matrix to detect statistically reliable aspects of the correlation matrix are seen as filtering procedures of the correlation matrix. We also discuss a method to associate a hierarchically nested factor model to a hierarchical tree obtained from a correlation matrix. The information retained in filtering procedures and its stability with respect to statistical fluctuations is quantified by using the Kullback-Leibler distance.   
\end{abstract}

\end{frontmatter}


\section{Introduction}
\label{intro}
Many complex systems observed in the physical, biological and social sciences are organized in a nested hierarchical structure, i.e. the elements of the system can be partitioned in clusters which in turn can be partitioned in subclusters and so on up to a certain level (Simon, 1962).  The hierarchical structure of interactions among elements strongly affects the dynamics of complex systems. Therefore a quantitative description of hierarchies of the system is a key step in the modeling of complex systems (Anderson, 1972). 
The analysis of multivariate data provides crucial information in the investigation of a wide variety of systems. Multivariate analysis methods are designed to extract the information both on the number of main factors characterizing the dynamics of the investigated system and on the composition of the groups (clusters) in which the system is intrinsically organized. Recently physicists started to contribute to the development of new techniques to investigate multivariate data (Blatt et al., 1996; Hutt et al., 1999; Mantegna, 1999; Giada and Marsili, 2001; Kraskov et al., 2005; Tumminello et al., 2005; Tsafrir et al., 2005; Slonim, 2005). Among multivariate techniques, natural candidates for detecting the hierarchical structure of a set of data are hierarchical clustering methods (Anderberg, 1973). 

The  modeling of the correlation matrix of a complex system with tools of hierarchical clustering has been useful in the multivariate characterization of stock return time series (Mantegna, 1999; Bonanno et al., 2001; Bonanno et al., 2003), market index returns 
of worldwide stock exchanges (Bonanno et al., 2000), and volatility increments of stock return time series (Micciche et al., 2003), where the estimation of statistical reliable properties of the correlation matrix is crucial for several financial decision processes such as asset allocation, portfolio optimization (Tola et al., 2008), derivative pricing, etc. We have termed the selection of statistical reliable information of the correlation matrix with the locution "filtering procedure" in Ref. Tumminello et al. (2007a). Hierarchical clustering procedures are filtering procedures. Other filtering procedures which are popular within the econophysics community are procedures based on the random matrix theory (Laloux et al., 1999; Plerou et al., 1999; Rosenow et al., 2002; Coronnello et al., 2005; Potters et al., 2005; Tumminello et al., 2007a), and procedures using the concept of shrinkage of a correlation matrix (Ledoit and Wolf, 2003; Sch\"afer and Strimmer, 2005; Tumminello et al., 2007b). Many others might be devised and their effectiveness tested. 

The correlation matrix of the time series of a multivariate complex system can be used to extract information about aspects of hierarchical organization of such a system. The clustering procedure is done by using the correlation between pairs of elements as a similarity measure and by applying a clustering algorithm to the correlation matrix. As a result of the clustering procedure, a hierarchical tree of the elements of the system is obtained. The correlation based clustering procedure allows also to associate  a correlation based network with the correlation matrix. For example, it is natural to select the minimum spanning tree, i.e. the shortest tree connecting all the elements in a graph, as the correlation based network associated with the single linkage cluster analysis. Different correlation based networks can be associated with the same hierarchical tree putting emphasis on different aspects of the sample correlation matrix. Useful examples of correlation based networks different from the minimum spanning tree are the planar maximally filtered graph (Tumminello et al., 2005) and the average linkage minimum spanning tree (Tumminello et al., 2007c). 

In correlation based hierarchical investigations the statistical reliability of hierarchical trees and networks is depending on the statistical reliability of the sample correlation matrix. The sample correlation matrix is computed by using a finite number of records $T$ sampling the behavior of the $N$ elements of the system. Due to the unavoidable finiteness of $T$, the estimation of the sample correlation matrix presents a degree of statistical uncertainty that can be characterized under widespread statistical assumptions. Physicists (Laloux et al., 1999; Plerou et al., 1999), have contributed to the quantitative estimation of the statistical uncertainty of the correlation matrix by using tools and concepts of random matrix theory. However, theoretical results providing the statistical reliability of hierarchical trees and correlation based networks are still not available and therefore, a bootstrap approach has been used to quantify the statistical reliability of both hierarchical trees (Tumminello et al., 2007d) and correlation based networks (Tumminello et al., 2007c).

The hierarchical tree characterizing a complex system can also be used to extract a factor model with independent factors acting on different elements in a nested way. In other words, the number of factors controlling each element may be different and different factors may act at different hierarchical levels. Tumminello et al. (2007d) have shown how to associate a hierarchically nested factor model to a system described by a given hierarchical structure. 

Having available a large number of filtering procedures, researchers encounter the necessity to have a quantitative methodology able to estimate the information retained in a filtered correlation matrix obtained from the sample correlation matrix. It is also important to quantify the stability of the filtering procedure in different realizations or replicas of the process and a distance of the filtered correlation matrix from a given reference model.
For all the above listed purposes, a very useful measure is the one using the Kullback-Leibler distance that was introduced in Tumminello et al. (2007a). This distance presents the important characteristics that its value quantifying the distance between a sample correlation matrix and the correlation matrix of the generating model turns out to be independent from the specific correlation matrix of the model both for multivariate Gaussian variables (Tumminello et al., 2007a) and for multivariate Student's $t$ variables (Biroli et al., 2007; Tumminello et al., 2007b).

In the present paper we discuss in a coherent and self-consistent way (i) some filtering procedures of the correlation matrix based on hierarchical clustering and the bootstrap validation of hierarchical trees and correlation based networks, (ii) the hierarchically nested factor model, (iii) the Kullback-Leibler distance between the probability density functions of two sets of multivariate random variables and (iv) the retained information and stability of a filtered correlation matrix. We apply the discussed concepts to a portfolio of stocks traded in a financial market. The paper is organized as follows. In Section \ref{corr} we discuss how to obtain hierarchical trees and correlation based trees or networks from the correlation matrix of a complex system and we discuss about the role of bootstrap in the statistical validation of hierarchical trees and correlation based networks. In Section \ref{sectionH} we discuss the definition and the properties of a factor model with independent factors which are hierarchically nested. In Section \ref{sectionEMP} we present an empirical application of the hierarchically nested factor model. Section \ref{kullbackTHEOR} discusses how to quantify the information and stability of a correlation matrix by using a Kullback-Leibler distance and Section \ref{sectionCOMP} presents the quantitative comparison of different filtering procedures performed with the same distance.  Section \ref{sectionCONC} briefly presents some conclusions.    

\section {Correlation based hierarchical organization and networks}
\label{corr}

Hereafter we discuss a simple example illustrating two filtering procedures of a correlation matrix performed with methods of hierarchical clustering. In our approach, by using the correlation between elements as the similarity measure and by applying a given hierarchical clustering procedure, we first obtain a hierarchical tree.  The information present in the hierarchical tree is completely equivalent to the information stored in the filtered matrix and, when the correlation is non-negative for each pair of elements, this matrix is positive definite (Tumminello et al., 2007d).

Our example considers the correlation matrix of $N=10$ daily stock returns traded at the New York Stock Exchange during the time period from January 2001 to December 2003. The investigated stocks are AIG, IBM, BAC, AXP, MER, TXN, SLB, MOT, RD, and OXY. The above presentation order of stocks is given according to their market capitalization at December 2003. In this paper, we indicate stocks with their tick symbols. The tick symbol, company name and other information of each company are given in Table I. From the Table we note that three stocks (OXY, RD, SLB) belong to the energy sector, three (IBM, MOT, TXN) to the technology sector and four (AIG, AXP, BAC, MER) to the financial sector.  

The stock return correlation matrix computed by using $T=748$ records is the following
\begin{equation}
\label{corrmat}
{\bf C}=\left(
\begin{array}{cccccccccc}
1.000	&	0.413	&	0.518	&	0.543	&	0.529	&	0.341	&	0.271	&	0.231	&	0.412	&	0.294\\
 ~	&	1.000	&	0.471	&	0.537	&	0.617	&	0.552	&	0.298	&	0.475	&	0.373	&	0.270\\
~	&	~	&	1.000	&	0.547	&	0.592	&	0.400	&	0.258	&	0.349	&	0.370	&	0.276\\
~	&	~	&	~	&	1.000	&	0.664	&	0.422	&	0.347	&	0.351	&	0.414	&	0.269\\
~	&	~	&	~	&	~	&	1.000	&	0.533	&	0.344	&	0.462	&	0.440	&	0.318\\
~	&	~	&	~	&	~	&	~	&	1.000	&	0.305	&	0.582	&	0.355	&	0.245\\
~	&	~	&	~	&	~	&	~	&	~	&	1.000	&	0.193	&	0.533	&	0.591\\
~	&	~	&	~	&	~	&	~	&	~	&	~	&	1.000	&	0.258	&	0.166\\
~	&	~	&	~	&	~	&	~	&	~	&	~	&	~	&	1.000	&	0.590\\
~	&	~	&	~	&	~	&	~	&	~	&	~	&	~	&	~	&	1.000\\
\end{array}
\right)
\end{equation}
where the order of elements of the correlation matrix from left to right and from top to bottom is the one based on capitalization given above.

A large number of hierarchical clustering procedures can be found in the literature. For a review about the classical techniques see, for instance, Anderberg (1973). In this paper we focus our attention on the single linkage cluster analysis (SLCA) and average linkage cluster analysis (ALCA).\\

The starting point of both the procedures is the empirical correlation matrix ${\bf C}$. The following procedure performs the ALCA giving as an output a hierarchical tree and a filtered correlation matrix ${\bf{C}}^<_{ALCA}$ 
:
\begin{enumerate} 

\item Set ${\bf{B}} = {\bf{C}}$.
\item Select the maximum correlation $b_{hk}$ in the correlation matrix ${\bf{B}}$. Note that $h$ and $k$ can be simple elements (i.e. clusters of one element each) or clusters (sets of elements). $\forall \, i \in h$ and $\forall \, j \in k$ one sets the elements $\rho^<_{ij}$ of the matrix ${\bf{C}}^<_{ALCA}$ as $\rho^<_{ij}=\rho^<_{ji}=b_{hk}$.
\item Merge cluster $h$ and cluster $k$ into a single cluster, say $q$. The merging operation identifies a node in the rooted tree connecting clusters $h$ and $k$ at the correlation $b_{hk}$. 
\item Redefine the matrix ${\bf{B}}$:
\begin{eqnarray} \label{negspin}
\left \{  \begin{aligned}
        &   b_{qj}= \frac{n_h\,b_{hj}+n_k \, b_{kj}}
                               {n_h+n_k}  & 
                           ~~~~\text{ if } j\notin h \,{\rm{and}} \, j\notin k\\
        &                 \nonumber \\
        &    b_{ij}=b_{ij} & 
                           ~~~~\text{ otherwise, }\\
\end{aligned} \right.
\end{eqnarray}
where $n_h$ and $n_k$ are the number of elements belonging respectively to the cluster $h$ and to the cluster $k$ before the merging operation. Note that if the dimension of ${\bf{B}}$ is $m \times m$ then the dimension of the redefined ${\bf{B}}$ is $(m-1) \times (m-1)$ because of the merging of clusters $h$ and $k$ into the cluster $q$.
\item If the dimension of ${\bf{B}}$ is larger than 1 then go to step (ii), else Stop.   
\end{enumerate}
By replacing point $(iv)$ of the above algorithm with the following item\\

\indent (iv)$'$  Redefine the matrix ${\bf{B}}$:
\begin{equation}\label{negspin2}
\left \{ \begin{aligned}
        &  b_{qj}= Max \left[b_{hj}, b_{kj}\right] & 
                           ~~~~\text{ if } j\notin h \,{\rm{and}} \, j\notin k \nonumber\\
        &  b_{ij}=b_{ij} & 
                           ~~~~\text{ otherwise, }\\
\end{aligned} \right.
\end{equation}
one obtains an algorithm performing the SLCA and the associated filtered correlation matrix  ${\bf{C}}^<_{SLCA}$. 

\begin{figure}
\includegraphics[width=1.0\textwidth]{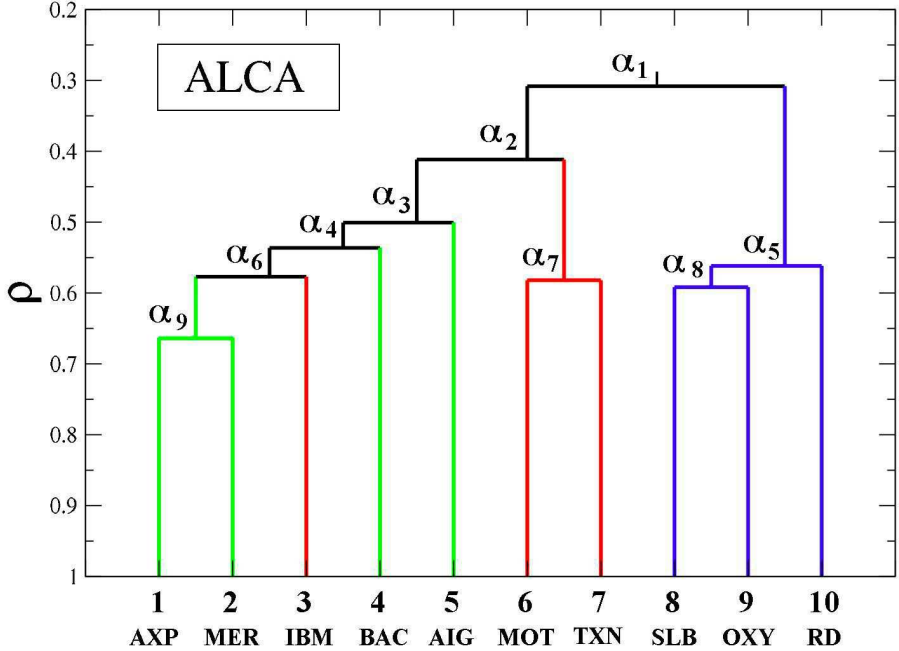}
\caption{ Average linkage cluster analysis. Illustrative example of a hierarchical tree associated to a system of $N=10$ stocks (tick symbols label stocks at the bottom of the hierarchical tree. Each element of the system is also labeled with an integer number). The color of line indicates the primary economic sector of the stock, red for technology, blue for energy and green for financial. The labels of the nodes of the hierarchical tree are used in the discussion of the hierarchically nested factor model of Section \ref{sectionH}.} 
\label{dendroALCA}
\end{figure}

\begin{figure}
\includegraphics[width=1.0\textwidth]{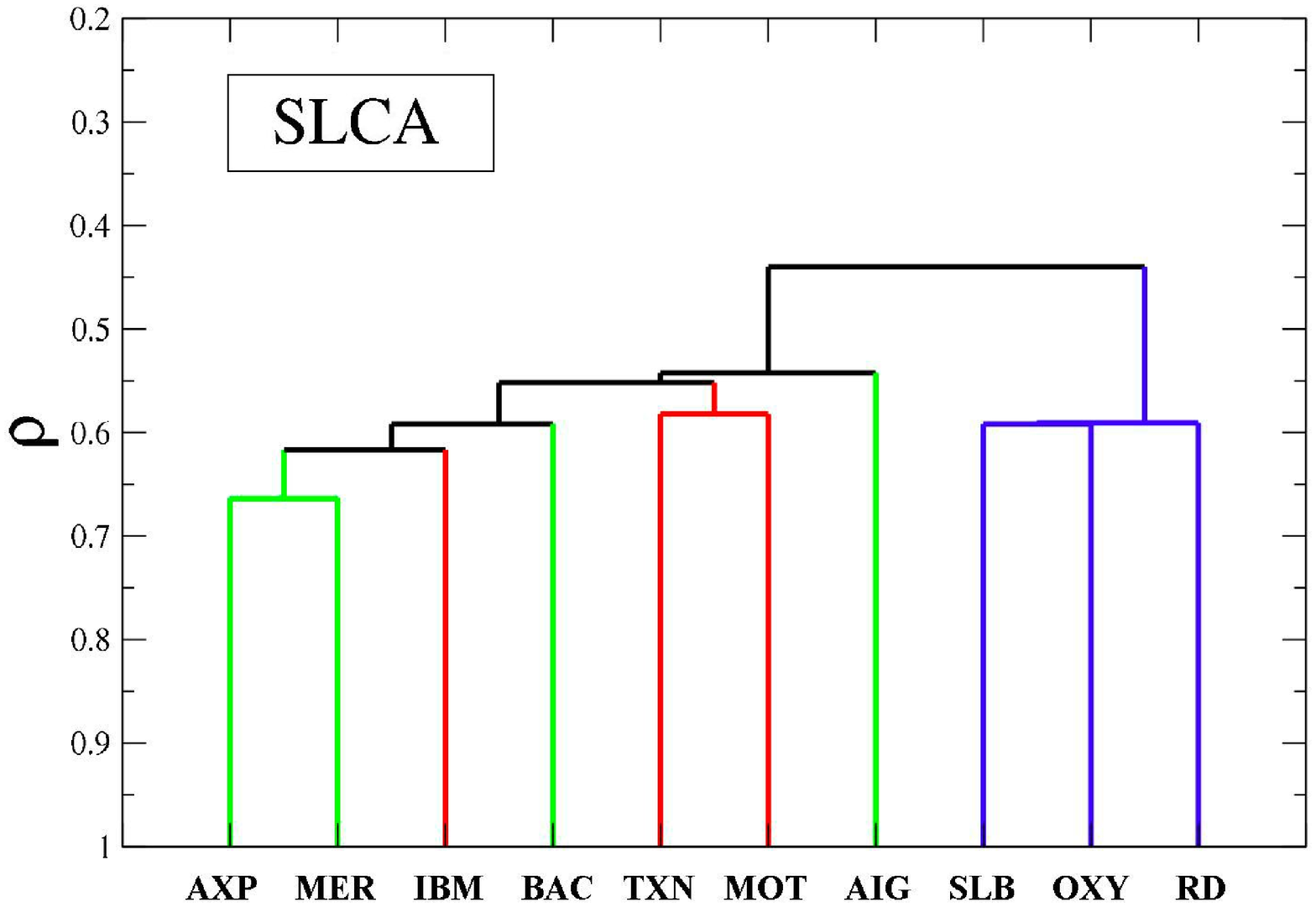}
\caption{Single linkage cluster analysis. Illustrative example of a hierarchical tree associated to a system of $N=10$ stocks (tick symbols label stocks at the bottom of the hierarchical tree). The color of line indicates the primary economic sector of the stock, red for technology, blue for energy and green for financial.} 
\label{dendroSLCA}
\end{figure}

The hierarchical trees obtained from the sample correlation matrix of Eq.~(\ref{corrmat}) by applying the ALCA and the SLCA are given in Fig. \ref{dendroALCA} and in  Fig. \ref{dendroSLCA} respectively. 
A hierarchical tree is a rooted tree, i.e. a tree in which a special node (the root) is singled out. In our example this node is $\alpha_1$.  In the rooted tree, we distinguish between leaves and internal nodes . Specifically, vertices of degree $1$ represent leaves (vertices labeled $1, 2,..., 10$ in Fig. \ref{dendroALCA}) while vertices of degree greater than 1 represent internal nodes (vertices labeled $\alpha_1$, $\alpha_2$,..., $\alpha_9$ in Fig. \ref{dendroALCA}). 
 The two trees are slightly different showing that each clustering method produce a different output putting emphasis on different aspects of the sample correlation matrix. 
 
The filtered correlation matrix associated with the ALCA is
\begin{equation}
\label{corrmatALCA}
{\bf{C}}^<_{ALCA}=\left(
\begin{array}{cccccccccc}
1.000	&	0.501	&	0.501	&	0.501	&	0.501	&	0.412	&	0.308	&	0.412	&	0.308	&	0.308\\
~	&	1.000	&	0.536	&	0.577	&	0.577	&	0.412	&	0.308	&	0.412	&	0.308	&	0.308\\
~	&	~	&	1.000	&	0.536	&	0.536	&	0.412	&	0.308	&	0.412	&	0.308	&	0.308\\
~	&	~	&	~	&	1.000	&	0.664	&	0.412	&	0.308	&	0.412	&	0.308	&	0.308\\
~	&	~	&	~	&	~	&	1.000	&	0.412	&	0.308	&	0.412	&	0.308	&	0.308\\
~	&	~	&	~	&	~	&	~	&	1.000	&	0.308	&	0.582	&	0.308	&	0.308\\
~	&	~	&	~	&	~	&	~	&	~	&	1.000	&	0.308	&	0.562	&	0.591\\
~	&	~	&	~	&	~	&	~	&	~	&	~	&	1.000	&	0.308	&	0.308\\
~	&	~	&	~	&	~	&	~	&	~	&	~	&	~	&	1.000	&	0.562\\
~	&	~	&	~	&	~	&	~	&	~	&	~	&	~	&	~	&	1.000\\
\end{array}
\right)
\end{equation}
%
whereas for the SLCA we obtain
%
\begin{equation}
\label{corrmatSLCA}
{\bf{C}}^<_{SLCA}=\left(
\begin{array}{cccccccccc}
1.000	&	0.543	&	0.543	&	0.543	&	0.543	&	0.543	&	0.440	&	0.543	&	0.440	&	0.440\\
~	&	1.000	&	0.592	&	0.617	&	0.617	&	0.552	&	0.440	&	0.552	&	0.440	&	0.440\\
~	&	~	&	1.000	&	0.592	&	0.592	&	0.552	&	0.440	&	0.552	&	0.440	&	0.440\\
~	&	~	&	~	&	1.000	&	0.664	&	0.552	&	0.440	&	0.552	&	0.440	&	0.440\\
~	&	~	&	~	&	~	&	1.000	&	0.552	&	0.440	&	0.552	&	0.440	&	0.440\\
~	&	~	&	~	&	~	&	~	&	1.000	&	0.440	&	0.582	&	0.440	&	0.440\\
~	&	~	&	~	&	~	&	~	&	~	&	1.000	&	0.440	&	0.590	&	0.591\\
~	&	~	&	~	&	~	&	~	&	~	&	~	&	1.000	&	0.440	&	0.440\\
~	&	~	&	~	&	~	&	~	&	~	&	~	&	~	&	1.000	&	0.590\\
~	&	~	&	~	&	~	&	~	&	~	&	~	&	~	&	~	&	1.000\\
\end{array}
\right)
\end{equation}
For the sake of comparison, here the ALCA and SLCA filtered correlation matrices are both written with the same order of stocks of the sample correlation matrix. By comparing the sample and the filtered matrices one immediately notes that the filtered ones contain less information being defined by a number of distinct correlation coefficients equals to $n-1$ whereas the original matrix has $n(n-1)/2$ distinct correlation coefficients. The two filtering methods detect different information. In fact the ALCA uses the average correlation coefficient between distinct groups of elements whereas the SLCA uses the maximal correlation. The two choices filter correlation coefficients characterized by a different degree of representativeness and statistical reliability.  

It is worth noting that the hierarchical methods reveal the sectorial structure of the considered set of stocks. Specifically, in both cases the stocks belonging to the energy sector form a cluster. Fig. \ref{dendroALCA} shows that for the ALCA dendrogram the node $\alpha_2$ splits the stocks in two sets, one composed by two technology stocks and one composed by the financial stocks plus the IBM. For the SLCA dendrogram the separation of the set in the technology and financial subsectors is less sharp (see Fig. \ref{dendroSLCA}). However in general hierarchical methods perform quite well in identifying groups of stocks belonging to the same economic sector (Mantegna, 1999; Bonanno et al., 2001; Coronnello et al., 2005).

In addition to the hierarchical trees and to the related filtered correlation matrices one can also obtain correlation based networks. 
Here we briefly recall how to select a correlation based graph out of the complete graph describing the system. A complete graph is a graph with links connecting all the elements (or nodes in the graph terminology) of the system of interest. In correlation based networks a weigth, which is monotonically related to the correlation coefficient of each pair of elements, can be associsted with each link. Therefore one can immediately associates a weighted completed graph with the correlation matrix among $n$ elements of interest. A complete graph is too rich of information and therefore a "filterin" (or "pruning") of it can improve its readability. For this reason a procedure can be set to select a subset of links which are highly informative about the hierarchical structure of the system. By using clastering algorithms as filtering procedures a certain number of correlation based graphs have been investigated in the econophysics literature. Correlation based networks which have been found very useful in the elucidation of economic properties of stock returns traded in a financial market are the minimum spanning tree (MST) (Mantegna, 1999), the planar maximally filtered graph (PMFG) (Tumminello et al., 2005) and the average linkage minimum spanning tree (ALMST) (Tumminello et al., 2007c). In the cited cases all the elements of the system are connected within the graph. Correlation based graphs with elements disconnected from a giant component can also be obtained starting from the correlation matrix. For  example, an extension from trees to more general graphs generated by selecting the most correlated links has been proposed in Onnela et al. (2003). However, this last method selects only a subset of the investigated elements controlled by an arbitrarely chosen threshold.

The MST is a correlation based tree associated with the SLCA. An illustrative algorithm providing the MST is the following.  Let us first recall that the connected component of a graph $g$ containing the vertex $i$ is the maximal set of vertices $S_i$ (with $i$ included) such that there exists a path in $g$ between all pairs of vertices belonging to $S_i$. When the element $i$ has no links to other vertices then $S_i$ reduces just to the element $i$. The starting point of the procedure is an empty graph $g$ with $N$ vertices. The MST algorithm can be summarized in 6 steps: 
\begin{enumerate}

\item Set ${\bf Q}$ as the matrix of elements $q_{ij}$ such that ${\bf Q}={\bf C}$. 

\item Select the maximum correlation $q_{hk}$ between elements belonging to different connected components  $S_h$ and $S_k$ in $g$. At the first step of the algorithm connected components coincide with single vertices in $g$. 

\item Find elements $u$, $p$ such that $\rho_{up}=\text{{\bf Max}}\left \{\rho_{ij}, \forall i \in S_h \text{ and } \forall j \in S_k \right \}$

\item Add to $g$ the link between elements $u$ and $p$ with weight $\rho_{up}$. Once the link is added to $g$, $u$ and $p$ will belong to the same connected component $S=S_h\bigcup S_k$. 

\item Redefine the matrix ${\bf Q}$: 

\begin{equation}\label{singleadjust}
\left\{
\begin{aligned}
q_{ij} & =q_{hk}, \text{ if } i\in S_h \text{ and } j\in S_k \\
q_{ij} & =\text{{\bf Max}} \left\{ q_{pt}, p\in S \text{ and } t\in S_j, \text{ with } S_j\neq S \right \}, \\
         & ~~~  \text{ if } i\in S \text{ and } j\in S_j \\
q_{ij} & =q_{ij}, \text{ otherwise}; \\
\end{aligned}
\right.
\end{equation} 
%

\item If $G$ is still a disconnected graph then go to step (ii), else stop.\\
\end{enumerate}

\begin{figure}
\includegraphics[width=1.0\textwidth]{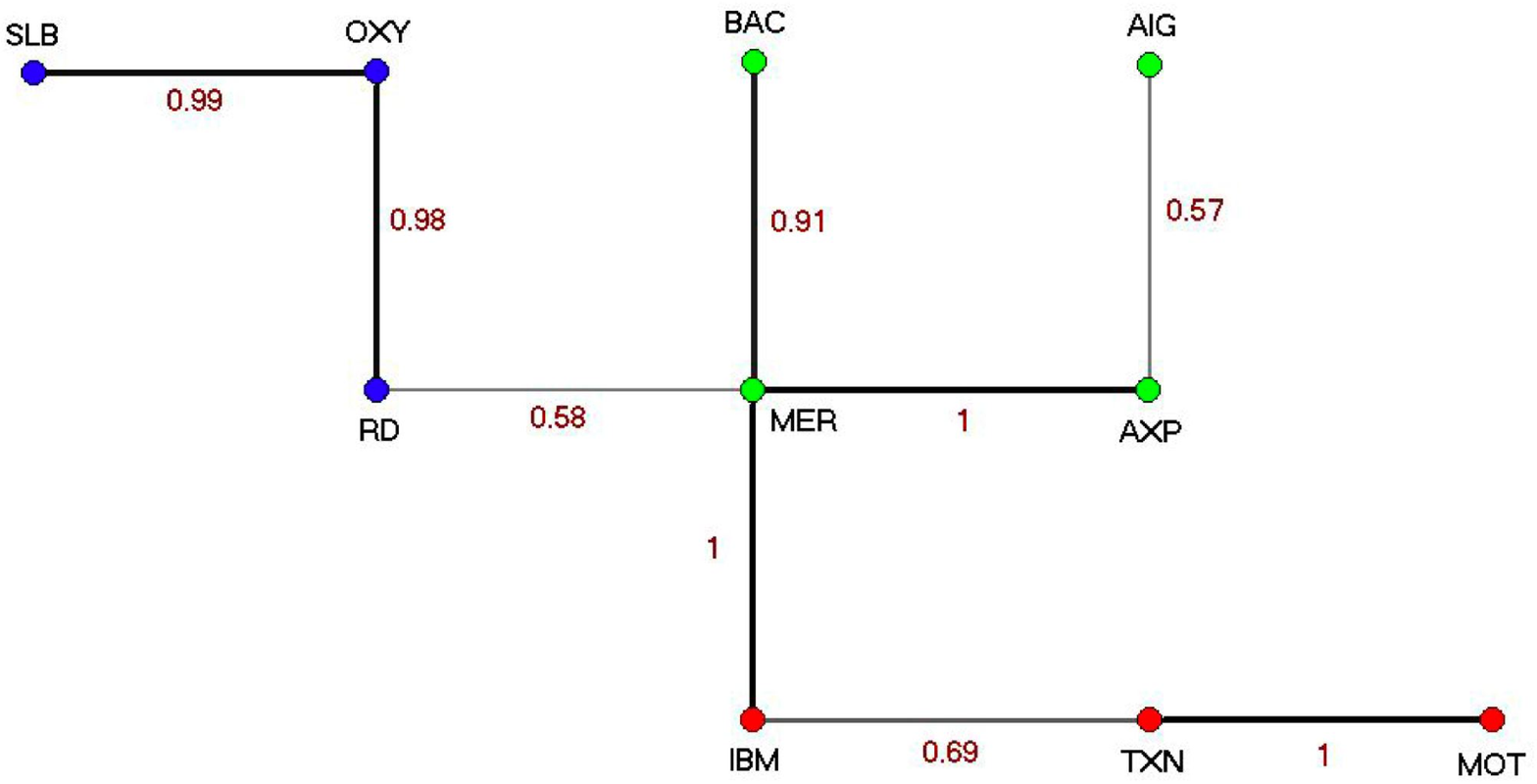}
\caption{Minimum spanning tree associated with the SLCA of the example. The vertices indicate the stocks. Colors indicate the different economic sectors, red for technology, blue for energy and green for financial. The thickness of links is proportional to the bootstrap percentage and this value is the number close to each link.} 
\label{mst10}
\end{figure}

The resulting graph $g$ is the MST of the system and the matrix ${\bf Q}$ is the correlation matrix associated to the SLCA. The presented algorithm is not the most popular or the simplest algorithm for the construction of the MST but it clearly reveals the relation between SLCA and MST. Indeed connected components progressively merging together during the construction of $g$ are nothing else but clusters progressively merging together in the SLCA.
In Fig. \ref{mst10} we show the MST associated with the considered example. It should be noted that the correlation based tree contains more information than the hierarchical tree or the filtered correlation matrix. For example, the fact that the connection between the cluster of two technology stocks (MOT and TXN) and the cluster of mostly financial stocks (MER, AXP, AIG, BAC and IBM) occurs through IBM is something which is not contained in the hierarchical tree but it is present in the MST.

By replacing eq. (\ref{singleadjust}) with
\begin{equation}\label{averageadjust}
\left\{
\begin{aligned}
& q_{ij}=q_{hk}, \text{ if } i\in S_h \text{ and } j\in S_k \\
& q_{ij}=\text{{\bf Mean}}  \left\{ q_{pt}, p\in S \text{ and } t\in S_j, \text{ with } S_j\neq S \right \}, \\
& ~~~~~~~  \text{ if } i\in S \text{ and } j\in S_j \\
& q_{ij}=q_{ij}, \text{ otherwise};
\end{aligned}
\right.
\end{equation} 
 in the step (v) of  the above procedure one obtains an algorithm performing the ALCA and the final ${\bf Q}$ of the procedure is the correspondent correlation matrix. The obtained tree $g$ that we termed ALMST (Tumminello et al., 2007c) is a tree naturally associated with such a clustering procedure. The choice of the link at step (iii) of the ALMST construction algorithm does not affect the clustering procedure but specify the construction of the correlation based tree. More precisely by selecting any link between nodes $u \in S_h \text{ and } p \in S_k $ the matrix ${\bf Q}$ representing the result of ALCA remains the same in terms of hierarchical tree. This degeneracy allows one to consider different rules to select the link between elements $u$ and $p$ at the step (iii) of the construction algorithm. Different rules at step (iii) give rise to different correlation based trees. The same observation holds true for the algorithm that generates the MST. This fact implies that in principle one can consider spanning trees which are different from the MST and are still associated with the SLCA. However, we have already recalled that the MST is unique in the sense that, when an Euclidean distance is defined between links of the spanning tree, MST is the spanning tree of shortest length (West, 2001). 

For the present example the ALMST is essentially indistinguishable from the MST and for this reason we will not display it here. It is worth noting that whereas the hierarchical trees obtained with ALCA and SLCA show slight differences, these differences essentially disappears at the level of the associated correlation based trees in the present example. 

Starting from the sample correlation matrix one can also obtain correlation based networks having a structure more complex than a tree.  One of such correlation based networks is the PMFG (Tumminello et al., 2005). This correlation based network has associated a hierarchical structure which is the one given by SLCA but it presents a graph structure which is richer than the one of the MST. In fact, the PMFG has loops and cliques. A clique of $k$ elements is a complete subgraph that links all $k$ elements. Due to topological constraints, only cliques of 3 and 4 elements are allowed in the PMFG.  To illustrate the PMFG algorithm, let us first consider a different construction algorithm for the MST. Following the ordered list  $S_{ord}$ of correlation coefficients starting from the couple of elements with largest correlation one adds a link between element $i$  and element $j$ if and only if the graph obtained after the link insertion is still a forest or it is a tree. A forest is a disconnected graph in which any two elements are connected by at most one path, i.e. a disconnected ensemble of trees. With this procedure, equivalent to the algorithm above detailed, the graph obtained after all links of $S_{ord}$ are considered is the MST.  In direct analogy, Tumminello et al. (2005) introduce a correlation based graph obtained by connecting elements with largest correlation under the topological constraint of fixed genus $G=0$. The genus is a topologically invariant property of a surface defined as the largest number of nonintersecting simple closed curves that can be drawn on the surface without separating it. Roughly speaking, it is the number of holes in a surface. The construction algorithm for such graph is: following the ordered list $S_{ord}$ starting from the couple of elements with largest correlation one adds a link between element $i$ and element $j$  if and only if the resulting graph can still be embedded on a plane or a sphere, i.e. topological surfaces with $G=0$. A basic difference of the PMFG with respect to the MST is the number of links which is $N-1$ in the MST and $3\,(N-2)$ in the PMFG. Moreover, the PMFG is a network with loops whereas the MST is a tree. It is worth recalling that Tumminello et al. (2005) have proven that the PMFG always contains the MST. 

\begin{figure}
\includegraphics[width=1.0\textwidth]{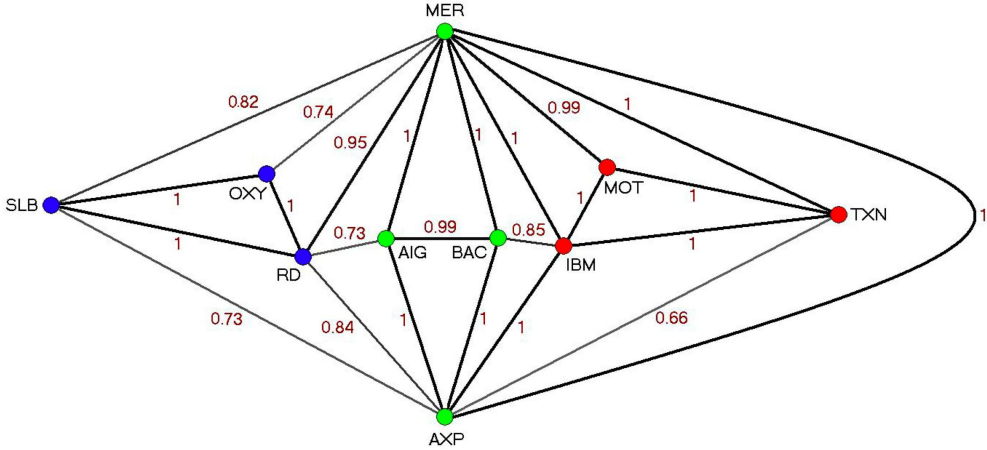}
\caption{Planar maximally filtered graph of the correlation matrix of the considered example. The vertices indicate the stocks. Colors indicate the different economic sectors, red for technology, blue for energy and green for financial. The thickness of links is proportional to the bootstrap percentage and this value is the number close to each link.} 
\label{pmfg10}
\end{figure}

In Fig. \ref {pmfg10} we show the PMFG obtained for the considered example. In the figure the length of the links is not related to the similarity measure between the two vertices they connect. We are using this kind of representation to put emphasis on the topological planarity of the network. In fact from the figure it is evident that there are no crossings of links, and the entire network is topologically embedded in a plane. By comparing Fig.s \ref{mst10} and  \ref{pmfg10} we note that the stock MER which turns out to be of central reference in the MST and ALMST is the only stock partecipating to all the seven 4-cliques which are observed in the PMFG. In other words, the PMFG allows to consider more details present in the sample correlation matrix than those selected by the MST or ALMST. For example the PMFG of Fig. \ref {pmfg10} shows that two stocks of the financial sector (AXP and MER) are connected to stocks of both the tecnology and energy sector. Such a property was not present in the MST (or ALMST) where only MER was linking two stocks of the two sectors (specifically RD and IBM). The PMFG is therefore showing more details on the interrelations present among stocks than the MST. The PMFG has been recently used to investigate stock return multivariate time series in references Tumminello et al. (2005), Coronnello et al. (2005) and Tumminello et al. (2007e).

The statistical reliability of regons of hierarchical trees and correlation based graphs cannot be theoretically evaluated in spite of the fact that the statistical reliability of the spectral properties of the correlation matrix can be assessed under the assumption of multivariate normal distribution for the time series of the elements of the investigated set. In the absence of such a theoretical approach we have devised a method to evaluate the statistical reliability of nodes in a hierarchical tree obtained by using a correlation matrix as a similarity measure and links in a correlation based graph. The method we use is based on a bootstrap procedure of the time series used to compute the correlation matrix of the system. The method is detailed in Tumminello et al. (2007d) and Tumminello et al. (2007c). Here we just sketch the most important aspects of the procedure allowing  to associate a bootstrap value to each internal node of a hierarchical tree. 
Consider a system of $N$ time series of length $T$ and suppose to collect data in a matrix ${\bf X}$ with $N$ columns and $T$ rows.  A bootstrap data matrix ${\bf X}^*$ is formed by randomly sampling $T$ rows from the original data matrix ${\bf X}$ allowing multiple sampling of the same row. For each replica ${\bf X}^*$, the associated correlation matrix ${\bf C}^*$ is evaluated and a hierarchical tree is constructed by hierarchical clustering. A large number (typically 1000) of independent bootstrap replicas is considered and for each internal node of the original data hierarchical tree we compute the fraction of bootstrap replicas (commonly referred to as bootstrap value) preserving the internal node in the hierarchical tree. Given an internal node $\alpha_k$ of the original hierarchical tree, we say that a bootstrap replica is preserving that node if and only if a node $\alpha_h^*$ in the replica hierarchical tree exists and identifies a branch characterized by the same leaves identified by $\alpha_k$ in the original hierarchical tree. For instance, we say that the node $\alpha_3$ of the hierarchical tree in Fig. \ref{dendroALCA} is preserved in some replica hierarchical tree $D^*$ if and only if a node of $D^*$ exists such that it connects all and only the leaves 1, 2, 3, 4, and 5.

\begin{figure}
\includegraphics[width=1.0\textwidth]{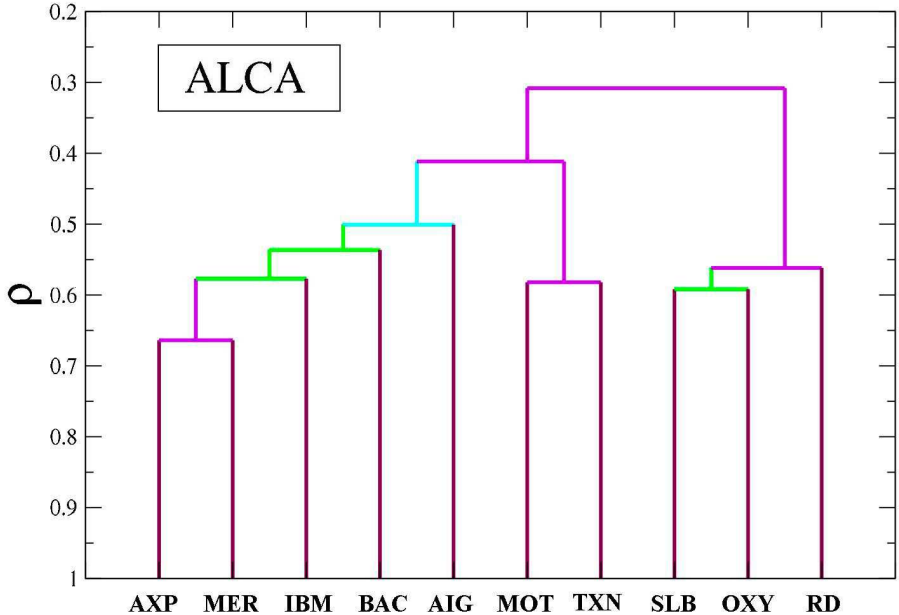}
\caption{Average linkage cluster analysis. Illustrative example of the estimation of the bootstrap value associated with each node of the hierarchical tree.  The color of horizontal lines indicate the bootstrap value $b$ of the node according to the following color code: Green  $0.4\le b < 0.6$, Cyan  $0.6\le b < 0.8$, and Purple  $0.8 \le b \le 1.0$.}  
\label{alcaBOOT}
\end{figure}

In Fig. \ref{alcaBOOT} we show the result of the application of the bootstrap procedure to the ALCA hierarchical tree of the example shown in Section \ref{corr}. From the figure it is evident that different nodes have a different statistical reliability as quantified through the bootstrap value. 

An analogous uncertainty is observed and quantified in correlation based graphs with a similar methodology. 
Consider a system of $N$ elements and suppose to collect data in a matrix ${\bf X}$ with $N$ columns and $T$ rows. A bootstrap data matrix ${\bf X}^*$ is formed by randomly sampling $T$ rows from the original data matrix ${\bf X}$ allowing multiple sampling of the same row. For each replica ${\bf X}^*$, the associated correlation matrix ${\bf C}^*$ is evaluated and a correlation based graph is constructed. By applying the procedures described in the previous subsection to ${\bf C}$ one can construct, for example, the MST, the PMFG and the ALMST of the system. The bootstrap technique requires to construct a number $r$ of replicas ${\bf X}^*_i$, $i\in{1,...,r}$ of the data matrix ${\bf X}$. Usually, $r=1000$ is considered a sufficient number of replicas. For each replica ${\bf X}^*_i$ the correlation matrix is evaluated and the correlation based graph of interest is obtained. The result is a collection of correlation based graphs. For example, in the case of MSTs, $\{MST^*_1,...,MST^*_{r} \}$ . To associate the so called \emph{bootstrap value} to a link of the original correlation based graph (in the present example a MST) one evaluates the number of $MST^*_i$ where the link is appearing and normalizes such a number with the total number of replicas, e.g. $r=1000$. The bootstrap value gives information about the reliability of each link of a correlation based graph. It is worth noting that the bootstrap approach does not require the knowledge of the data distribution and then it is particularly useful to deal with high dimensional systems where it is difficult to infer the joint probability distribution from data.  One might then be tempted to expect that the higher is the correlation associated to a link in a correlation based network the higher is the reliability of the link. Tumminello et al. (2007c) show that such hypothesis is not always observed in empirical results of sets of stock returns traded in a financial market. The bootstrap vaue and the correlation coefficient can be different indicating a different degree of stability with respect to metric and topological aspects. In Figs \ref{mst10} and \ref{pmfg10} the bootstrap values associated with each link are reported in the figure as the number close to each link. For a detailed discussion about the use of the bootstrap procedure to estimate the reliability of correlation based graphs see Tumminello et al. (2007c).

\section{A hierarchically nested factor model}
\label{sectionH}

The filtering procedure of the correlation matrix provides filtered correlation matrices carrying information on the hierarchical structure of the investigated system. A hierarchical factor model associated with such a matrix is useful in the modeling of hierarchical complex systems. In this Section we discuss the hierarchically nested factor model (HNFM). The HNFM is introduced in Tumminello et al. (2007d) in such a way that its correlation matrix coincides with the similarity matrix ${\bf C}^{<}$ filtered by a chosen hierarchical clustering procedure. 

Hereafter, we illustrate the methodology to associate a nested factor model to a multivariate data set. The association is done by retaining all the information about the hierarchies detected by a hierarchical clustering. This is achieved by considering a factor model in bijective relation with a hierarchical tree. We are going to present our method by making use of the illustrative hierarchical tree given in Fig. \ref{dendroALCA}.
We first note that to each leaf $i$ (or internal node $\alpha_h$) one can associate a \emph{genealogy}  $G(i)$  ($G(\alpha_h)$ for the internal node $\alpha_h$) which is the ordered set of internal nodes connecting leaf $i$ (internal node $\alpha_h$) to the root $\alpha_1$.
For instance the genealogy associated to the leaf 3 in the figure is $G(3)=\{\alpha_6, \alpha_4, \alpha_3, \alpha_2, \alpha_1 \}$ and the genealogy of the internal node $\alpha_7$ in the figure is $G(\alpha_7)=\{\alpha_7, \alpha_2, \alpha_1 \}$. 
Note that the internal node $\alpha_7$ is included in $G(\alpha_7)$. We say that an internal node $w$ is the {\it  parent} of the node $v$ and we use the notation $w=g(v)$ if $w$ immediately precedes $v$ on the path from the root to $v$. For example $\alpha_2=g(\alpha_7)$ in Fig. \ref{dendroALCA}.
Beside the topological structure, hierarchical trees obtained through clustering algorithms of the correlation matrix have also metric properties. In fact clustering algorithms associate to each internal node $\alpha_i$ a correlation coefficient  $\rho_{\alpha_i}$. Our internal node labeling implies that  $\rho_{\alpha_i} \le \rho_{\alpha_{i+1}}$ and here we consider $\rho_{\alpha_1}\ge0$ . 
In ${\bf C}^{<}$ there are at most $N-1$ distinct coefficients (see discussion in Section \ref{corr}). Exactly $N-1$ distinct coefficients are obtained in case of binary rooted trees. 

In Tumminello et al. (2007d) we introduce the factor model
\begin{equation}
\label{model}
x_i (t)=\sum_{\alpha_h \in G(i)} {\gamma_{\alpha_h} f^{(\alpha_h)}(t)}+\eta_i\,\, \epsilon_i (t)
\end{equation}
where $i\in\{1,...,N\}$, $\eta_i=[1-\sum_{\alpha_h \in G(i)} {\gamma_{\alpha_h}^{2}}]^{1/2}$,  the $h^{th}$ factor $f^{(\alpha_h)}(t)$ and $\epsilon_i$ are i.i.d. random variables with zero mean and unit variance. By fixing the $\gamma$ parameters as 
\begin{eqnarray}
\label{coeffic}
\gamma_{\alpha_1} = & \sqrt{\rho_{\alpha_1}} & \nonumber \\
\gamma_{\alpha_h} =  & \sqrt{\rho_{\alpha_h}-\rho_{g(\alpha_h)}} & \,\,\,\,\,\,\, \forall \, h=2,...,n-1
\end{eqnarray}
the model of Eq. (\ref{model}) is the factor model characterized by a correlation matrix equals to a given matrix ${\bf C}^<$. It should be noted that by assuming $\rho_{\alpha_1}\ge0$, all the coefficients $\gamma_{\alpha_h}$ are non negative real numbers. 
In Tumminello et al. (2007d) we prove that correlation $\rho_{i j}^< = \rho_{\alpha_k}$. In fact, the cross correlation 
\footnote{Here and in the following we indicate with the notation $\langle x \rangle$ the mean value of the $x$ variable.}
$\langle x_i x_j \rangle$ only depends on the factors $f^{(\alpha_h)}$ which are common to $x_i$ and $x_j$. Since one associates a factor to each internal node, one needs to identify the internal nodes belonging to both the genealogies $G(i)$ and $G(j)$. 
One can verifies that  $G(i) \cap G(j)=G(\alpha_k)$. For example, in Fig. \ref{dendroALCA} we have that  $G(5)=\{\alpha_3, \alpha_2, \alpha_1 \}$ and $G(6)=\{\alpha_7, \alpha_2, \alpha_1 \}$ so that $G(5)\cap G(6)=\{\alpha_2, \alpha_1 \}=G(\alpha_2)$. 
By making use of Eqs.~(\ref{model}, \ref{coeffic}) the cross correlation between variables $x_i$ and $x_j$ is
\begin{equation}
\label{scalar}
\left\langle x_i x_j \right\rangle = \sum_{\alpha_h \in G(\alpha_k)} {\gamma_{\alpha_h}^2} = \rho_{\alpha_k}=\rho_{i j}^<.
\end{equation} 
For example with reference to Fig.~\ref{dendroALCA} we have $\left\langle x_5 x_6 \right\rangle= \gamma_{\alpha_2}^2+\gamma_{\alpha_1}^2=\rho_{\alpha_2}-\rho_{\alpha_1}+\rho_{\alpha_1}=\rho_{\alpha_2}$.
Thus the matrix ${\bf C}^<$ is the correlation matrix associated with the factor model of Eq. (\ref{model}). It is worth noting that the existence of a factor model whose matrix ${\bf C}^<$ is the correlation matrix implies that the matrix ${\bf C}^<$ is always positive definite if $\rho_{\alpha_1}\ge0$.

In the case in which negative correlations are associated with some nodes in the dendrogram, it is sometimes 
possible to suitably modify Eqs. (\ref{coeffic}) by introducing multiplicative sign variables in order to get an
HNFM describing the system. The description of the most general case is left for a future work. Here we just 
consider the case in which only $\rho_{\alpha_1}<0$, because this is the case in the empirical application 
described in section \ref{sectionEMP}. Let us assume that all the correlations associated with nodes in the dendrogram 
are non negative but  $\rho_{\alpha_1}<0$. Furthermore assume that $\vert \rho_{\alpha_1}\vert<\rho_{\alpha_2}$ (this constraint is satisfied in the empirical application of Section \ref{sectionEMP}). In order to 
construct the HNFM, we divide the elements of the system into two groups. These are the two groups of elements 
merging together at root node. The coefficient $\gamma_{\alpha_1}$ shall be different for elements belonging to 
different groups. Specifically,

\begin{eqnarray}
\label{coefficalpha11}
\gamma^1_{\alpha_1} = & - \sqrt{\vert \rho_{\alpha_1} \vert} & \text{for all the elements of the first group} 
\nonumber \\
\gamma^2_{\alpha_1} = & \sqrt{\vert \rho_{\alpha_1} \vert} & \text{for all the elements of the second group}, 
\end{eqnarray}
whereas we don't need to distinguish among elements belonging to different groups for the other $\gamma$ 
coefficients. Specifically

\begin{eqnarray}
\label{coefficalpha22}
\gamma_{\alpha_h} =  & \sqrt{\rho_{\alpha_h}-\vert \rho_{g(\alpha_h)}\vert} & \,\,\,\,\,\,\, \forall \, h=2,...,n-1. 
\end{eqnarray}

We note that the constraint $\vert \rho_{\alpha_1}\vert<\rho_{\alpha_2}$ is required by Eq. (\ref{coefficalpha22}) in order to have a real value of $\gamma$ coefficients. 
It is also to notice that $\vert \rho_{g(\alpha_h)}\vert=\rho_{g(\alpha_h)}\, \forall \,g(\alpha_h)\ne\alpha_1$, and 
accordingly, the $\gamma$ coefficients associated with all of the nodes different from the root node and its sons 
as given in Eq. (\ref{coefficalpha22}) coincide with the corresponding coefficients as defined in Eq. (\ref{coeffic}).


Eq. (\ref{model}) defines a HNFM of $N-1$ factors obtained from a hierarchical tree of $N$ elements. In general the number of factors determining the dynamics of the system can be significantly smaller than $N-1$. Moreover a correlation coefficient matrix obtained from a finite multivariate time series has associated an unavoidable statistical uncertainty that might introduce spurious factors. 
To overcome this problem, in Tumminello et al. (2007d) we propose a method devised to select the HNFM characterized by the largest number of factors (although in any case less than $N$) compatible with a predefined threshold of statistical reliability of retained factors. The method of Tumminello et al. (2007d) exploits the technique of non parametric bootstrap (Efron, 1979; Efron and Tibshirani, 1994). 
The bootstrap technique allows to associate a bootstrap value to each internal node of a hierarchical tree. Due to the one by one relation between nodes in the hierarchical tree and factors in the HNFM, the bootstrap value associated to a certain node of the hierarchical tree is associated also to the corresponding factor in the HNFM. 

Since the bootstrap value is a measure of the node's reliability, we propose to remove those nodes, and therefore the corresponding factors, with bootstrap value smaller than a given threshold $b$. This is done by merging each node with a bootstrap value smaller than $b$ with its first ancestor node in the path to the root having a bootstrap value greater than $b$ and then constructing the HNFM associated with this reduced hierarchical tree. The question is how to select the threshold $b$. The bootstrap value of a certain node (factor) cannot be straightforwardly intended as the probability that the node (factor) belongs to the true and unknown hierarchy (model) of the system.
For example, in phylogenetic analysis Hillis and Bull (1993) have shown that bootstrap values of more than $70\%$ correspond to a probability of more than $95\%$ that the true phylogeny has been found.   In Tumminello et al. (2007d) we do not choose {\it a priori} the value of $b$ but we infer  a suitable value of the threshold from the data in a self consistent way.  
The detailed procedure used to determine the threshold from data is discussed in Tumminello et al. (2007d).

\section{An empirical application: a set of stocks traded in a financial market}
\label{sectionEMP}

\begin{figure}
\includegraphics[width=1.0\textwidth]{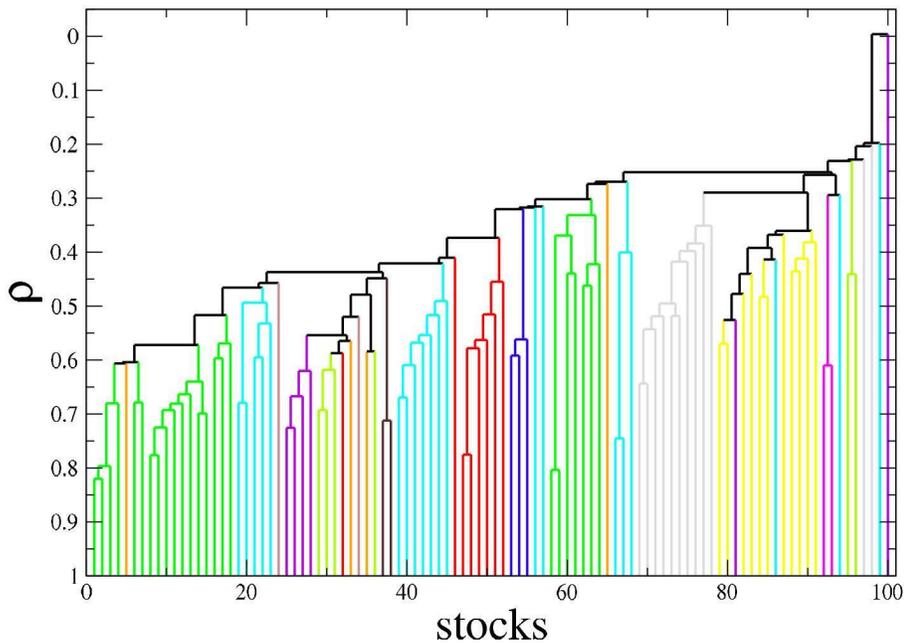}
\caption{Hierarchical tree of the set of daily equity return of $100$ highly capitalized stocks traded at the NYSE during the period 2001-2003 obtained by applying the average linkage clustering algorithm to the correlation matrix. Colors are chosen according to the stock economic sector according to the classification of Yahoo Finance. Specifically these sectors are Basic Materials (violet), Consumer Cyclical (tan), Consumer Non Cyclical (yellow), Energy (blue), Services (cyan), Financial (green), Healthcare (gray), Technology (red), Utilities (magenta), Transportation (brown), Conglomerates (orange) and Capital Goods (light green).} 
\label{dendroData}
\end{figure}

As an application of the described technique to real data we examine the set of daily equity return of $N=100$ highly capitalized stocks traded at the NYSE during the period 2001-2003 ($T=748$). Specifically, we apply the ALCA to the correlation matrix of the system and we obtain the hierarchical tree shown in Fig. \ref{dendroData}.  In the figure, the identity of each stock is labeled by an integer number. The correspondence between each number and the tick symbol of the stock is provided in Table I. In the same Table we also provide information about the company name and economic sector and sub-sector classified according to Yahoo Finance.

\begin{figure}[b,t]
\includegraphics[width=1.0\textwidth]{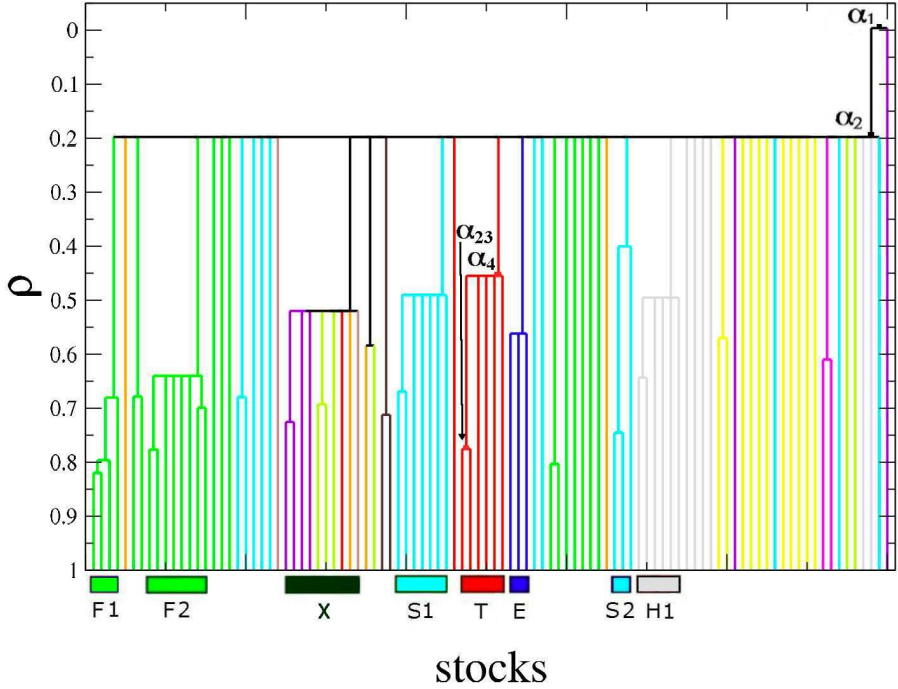}
\caption{Hierarchical tree with 27 internal nodes obtained by node reduction of the ALCA hierarchical tree shown in Fig. \ref{dendroData}. Rectangles at the bottom are indicating 8 clusters and the associated symbols label the classification of stocks in terms of economic sectors or sub-sectors according to the classification of Yahoo Finance (see text for the legend). Colors of lines indicate stock sectors as in Fig. \ref{dendroData}. The labeled internal nodes are discussed in the text. In the figure we do not comment on clusters composed by only two leaves.} 
\label{nodered}
\end{figure}

To evaluate the statistical robustness of each node and to simplify the description in terms of a HNFM we use the bootstrap technique discussed above. In particular, we select in a self-consistent way (see Tumminello et al. (2007d)) the bootstrap value threshold,
which turns out to be
$b=70\%$  for the considered dataset. 
The corresponding reduced hierarchical tree has 27 nodes and it is reported in Fig. \ref{nodered}. 

Let us first comment the properties of the reduced HNFM. In the figure we observe several clusters and sub-clusters. As already noticed in previous studies (Mantegna, 1999; Bonanno et al., 2001; Tumminello et al., 2005), the detected clusters and sub-clusters are overlapping in part with economic classification such as, for example, the one provided by the Yahoo Finance (at April 2005). This can be seen in Fig. \ref{dendroData} and \ref{nodered} where we use this classification to characterize with a specific color each stock. Most of the groups detected by hierarchical clustering are characterized by the same color. For example, financial firms are represented in Fig. \ref{dendroData} and \ref{nodered} as green lines in the hierarchical tree. 

The root of the dendrogram of Fig. \ref{nodered} is associated with a parameter $\rho_{\alpha_1}=-0.004$. This value is negative even if it is not statistically significantly different from zero given that the error associated with the correlation coefficient is $1/\sqrt{T}=0.036$. The fact that $\rho_{\alpha_1}<0$ requires the introduction of sign variables as explained at the end of Section \ref{sectionH}. Since it is $\rho_{\alpha_2}=0.2>|\rho_{\alpha_1}|$, we can use the Eq.s \ref{coefficalpha11} and \ref{coefficalpha22} to determine the parameters of the model. The root node $\alpha_1$ splits the set of stocks in two subsets, one composed by one stock (NEM, a gold mining company, see Table I) and another composed by $99$ stocks. The value of $\rho_{\alpha_1}$ is consistent with the interpretation that NEM is uncorrelated to the rest of the stocks. By using Eq. \ref{coefficalpha11} we set 
$\gamma^1_{\alpha_1}=-0.063$ and $\gamma^2_{\alpha_1}=+0.063$. The second factor (node) $\alpha_2$ describes the market mean behavior and it is associated with the parameter $\gamma_{\alpha_2}=0.44$. The other $25$ factors describe clusters of stocks that are often significantly homogeneous with respect to the sector activity of the stocks.   In Fig. \ref{nodered} we have highlighted 8 clusters by using rectangles at the bottom of the figure. Specifically, F1 is the sub-sector of {\it investment services} and F2 contains the sub-sectors of {\it regional banks} and {\it money center banks}. Both F1 and F2 belong to the economic sector of {\it Financial}; T and E are indicating the economic sectors of {\it Technology}  and {\it Energy} respectively; H1 indicates the sub-sector  {\it major drugs} of the economic sector {\it Healthcare}; S1 and S2 indicate the two sub-sectors of {\it retail} and {\it communication services} of the sector of {\it Services} respectively. Finally, X is a cluster which is not homogeneous with respect to sector and sub-sector classification. It comprises stocks in the sector of {\it basic materials}, stocks of the sub-sector {\it constructions} of {\it capital goods} and stocks as EMR (classified as {\it technology}) and GM  (classified as {\it consumer cyclical}).

One prominent example is the group of technology stocks (group T in Fig. \ref{nodered}).  The first two stocks (their tick symbols are TXN and ADI) from left to right of the group labeled as T in the reduced HNFM of Fig. \ref{nodered} are described by the equation 
\begin{eqnarray}
x_i^F(t)=\gamma_{\alpha_{23}} f^{(\alpha_{23})}(t)+\gamma_{\alpha_{4}} f^{(\alpha_{4})}(t)+\sum_{h=1}^{2}{\gamma_{\alpha_{h}} f^{(\alpha_h)}(t)} +\eta_F \epsilon_i(t)=\\
0.57f^{(\alpha_{23})}(t)+0.51 f^{(\alpha_{4})}(t)+0.44 f^{(\alpha_{2})}(t) +0.063 f^{(\alpha_{1})}(t)+0.47 \epsilon_i(t) \nonumber
\end{eqnarray}
The factors $f^{(\alpha_1)}(t)$ and $f^{(\alpha_2)}(t)$ are common to almost all stocks whereas $f^{(\alpha_4)}(t)$ and $f^{(\alpha_{23}}(t)$ are specific to these stocks. The other four technology stocks  (which are EMC, IBM, MOT and CA) are described by the equation
\begin{eqnarray}
x_i^F(t)=\gamma_{\alpha_{4}} f^{(\alpha_{4})}(t)+\sum_{h=1}^{2}{\gamma_{\alpha_{h}} f^{(\alpha_h)}(t)} +\eta_F \epsilon_i(t)=\\
0.51 f^{(\alpha_{4})}(t)+0.44 f^{(\alpha_{2})}(t) +0.063 f^{(\alpha_{1})}(t)+0.74 \epsilon_i(t) \nonumber.
\end{eqnarray}
In this last case only the $f^{(\alpha_4)}(t)$ factor is present in addition to the $f^{(\alpha_1)}(t)$ and $f^{(\alpha_2)}(t)$ factors common to almost all stocks. It is therefore natural to consider $f^{(\alpha_4)}(t)$ as a factor characterizing technology stocks whereas $f^{(\alpha_{23})}(t)$ is an additional factor further characterizing only the two stocks TXN and ADI. A similar organization in nested clusters is observed in all the groups detected by the reduced HNFM. The number of factors characterizing the various stocks is ranging from one to five.

It is worth to compare Fig. \ref{dendroData} and \ref{nodered}. The comparison shows that the self-consistent reduction of the number of factors allow a robust statistical validation of the groups that are detected from the data analysis. Only the information which is statistically robust at the $95\%$ level is retained in the reduced HNFM. For example, the financial cluster observed at the left end of the hierarchical tree in Fig. \ref{dendroData} is not robust at the selected confidence level whereas the two sub-clusters indicated as F1 (LEH, BSC, MER and SCH) and F2 (NCC, STI, ONE, PNC, BAC, WFC, BK and MEL) in Fig.\ref{nodered} are.
This empirical analysis has shown the usefulness of HNFM in an empirical investigation of hierarchically organized complex systems.
%




\section{Information and stability of a correlation matrix via the Kullback-Leibler distance}
\label{kullbackTHEOR}

In Tumminello et al. (2007a) we propose to measure the performance of filtering procedures by using the Kullback-Leibler distance introduced by Kullback and Leibler (1951). The Kullback-Leibler distance (see, for instance, Cover and Thomas (1991)) or \emph{mutual entropy} is a measure of the distance between two probability densities, say $p$ and $q$. It is defined as
\begin{equation}
\label{kullbackEQGEN}
K(p,q)=E_p\left[\log{\left( \frac{p}{q} \right)}\right],
\end{equation}
where $E_p[\, . \, ]$ indicates the expectation value with respect to the probability density $p$. The Kullback-Leibler distance is asymmetric. In Eq.(\ref{kullbackEQGEN}) the expectation value is evaluated according to the distribution $p$. 

We consider first the Kullback-Leibler distance between multivariate Gaussian random variables (Tumminello et al., 2007a). We consider variables with zero mean and unit variance without loss of generality because we are interested in the comparison of the correlation matrices of the two sets of variables. In this case, the Gaussian multivariate distribution associated with the random vector $X$ is completely defined by the correlation matrix ${\bf \Sigma}$ of the system. In the following we indicate the probability density function with $P({\bf \Sigma},X)$.
Given two different probability density functions $P({\bf \Sigma}_1,X)$ and $P({\bf \Sigma}_2,X)$, we have
%
 %
\begin{eqnarray}
\label{kullbackMulti}
K(P({\bf \Sigma}_1,X),P({\bf \Sigma}_2,X))=E_{P({\bf \Sigma}_1,X)}\left[\log{\left( \frac{P({\bf \Sigma}_1,X)}{P({\bf \Sigma}_2,X)} \right)}\right] = 
\nonumber \\
=\int{P({\bf \Sigma}_1,X) \log\left[ \frac{P({\bf \Sigma}_1,X)}{P({\bf \Sigma}_2,X)}\right] dX},
\end{eqnarray}
%
By performing the integral in Eq. (\ref{kullbackMulti}) one obtains:
\begin{eqnarray}
\label{kullbackGaussian}
K(P({\bf \Sigma}_1,X),P({\bf \Sigma}_2,X))=\frac{1}{2} \left[\log{\left(\frac{\abs{{\bf \Sigma}_2}}{\abs{{\bf \Sigma}_1}}\right)}+\text{tr}\left({{\bf \Sigma}_2^{-1} {\bf \Sigma}_1 }\right)-N\right],
\end{eqnarray}
where $N$ is the dimension of the space spanned by the $X$ variable and ${\abs{{\bf \Sigma}}}$ indicates the determinant of ${\bf \Sigma}$.
From now on we indicate $K(P({\bf \Sigma}_1,X),P({\bf \Sigma}_2,X))$ simply with $K({\bf \Sigma}_1,{\bf \Sigma}_2)$. It is worth noting that the Kullback-Leibler distance takes naturally into account the statistical nature of correlation matrices. Indeed $K({\bf \Sigma}_1,{\bf \Sigma}_2)$ is well defined only provided that the matrices ${\bf \Sigma}_1$ and ${\bf \Sigma}_2$ are positive definite. This property is not common to other measures of distance between matrices. However this property can also be a limitation. The Kullback-Leibler distance cannot be used to quantify the distance between semi-positive correlation matrices that are observed, for example, when  the length $T$ of data series is smaller than the number $N$ of elements of the system.

The Kullback-Leibler distance is also related to the maximum likelihood factor analysis (Mardia et al., 1979). In fact, the log-likelihood function to be maximized in order to describe a system of $N$ elements with sample correlation matrix ${\bf C}$ is a function of the Kullback-Leibler distance between the ${\bf C}$ and the model correlation matrix (see  Tumminello et al. (2007a) for details).

%
%


We have obtained the value of the Kullback-Leibler distance between two multivariate distribution as a function of the two corresponding Pearson correlation matrices. We are interested to  the case in which one or both correlation matrices are sample correlation matrices and thus are random variables.  Since different realizations of the process give rise to different sample correlation matrices, a Kullback-Leibler distance having one or two sample correlation matrices as arguments is a function of one or two random matrices.

In the case of multinormally distributed variables, we consider a random vector $X$ of dimension $N$ with a model correlation matrix ${\bf \Sigma}$. Let ${\bf C}_1$ and ${\bf C}_2$ be two sample correlation matrices obtained from two independent realizations of the system both of length $T$.
It is known that in this case sample covariance matrices belong to the ensemble of Wishart random matrices and many statistical properties of Wishart matrices are known (Mardia et al., 1979)). By making use of the theory of Wishart matrices, we obtained (Tumminello et al., 2007a) that
\begin{eqnarray}
\label{kullexpecSigS1}
E\left[K({\bf \Sigma},{\bf C}_1)\right]=\frac{1}{2} \left \{N\log{\left(\frac{2}{T}\right)}+
\sum_{p=T-N+1}^{T}{\left[\frac{\Gamma^{\prime}(p/2)}{\Gamma(p/2)}\right]}+\frac{N (N+1)}{T-N-1}\right\},
\end{eqnarray}
\begin{eqnarray}
\label{kullexpecS1Sig}
E\left[K({\bf C}_1,{\bf \Sigma})\right]=\frac{1}{2} \left \{N\log{\left(\frac{T}{2}\right)}
 -\sum_{p=T-N+1}^{T}{\left[\frac{\Gamma^{\prime}(p/2)}{\Gamma(p/2)}\right]}\right\}
\end{eqnarray}
and
 \begin{equation}
\label{kullexpecS1S2}
E\left[K({\bf C}_1,{\bf C}_2)\right]=\frac{1}{2} \frac{N (N+1)}{T-N-1},
\end{equation}
where 
%
$\Gamma(x)$ is the usual Gamma function and $\Gamma^{\prime}(x)$ is the derivative of $\Gamma(x)$. 
%
%

It is important to observe that all the expectation values given in Eq.s (\ref{kullexpecSigS1}-\ref{kullexpecS1S2}) are independent of ${\bf \Sigma}$, i.e. they are independent of the specific model generating or describing the data. The independence property implies that (i) the Kullback-Leibler distance is a good measure of the statistical uncertainty of correlation matrix which is due to the finite length of data series and (ii) the expected value of the Kullback-Leibler distance is known also when the underlying model hypothesized to describe the system is unknown. This fact has important consequences. Suppose one 
knows that the observed data are well approximated by a multivariate Gaussian distribution and that one measures a sample correlation matrix ${\bf C}$. In order to remove some unavoidably present statistical uncertainty, the observer applies a filtering procedure to the data obtaining the filtered
\footnote{In this and in the following sections we use the superscript to indicate the filtering procedure, whereas in Section 2 we used the subscript.} correlation matrix ${\bf C}^{filt}$. If the filtering technique is able to recover the model correlation matrix, i.e. ${\bf C}^{filt}={\bf \Sigma}$, the Kullback-Leibler distance $K({\bf C},{\bf C}^{filt})$ must be equal on average to the value given in Eq.~(\ref{kullexpecS1Sig}). This expected value is independent on the (unknown) model correlation matrix ${\bf \Sigma}$.   
Therefore large deviations from this expectation value indicate that the filtered matrix is not consistent with the true matrix of the system. If $K({\bf C},{\bf C}^{filt})$ is significantly smaller than the expectation value of Eq. (\ref{kullexpecS1Sig}) the filtered matrix is keeping some of the statistical uncertainty due to the finite length $T$. If, on the other hand, $K({\bf C},{\bf C}^{filt})$ is significantly larger than the value of Eq. (\ref{kullexpecS1Sig}), it means that the filtered matrix is either filtering too much information or distorting the signal. The distance between $K({\bf C},{\bf C}^{filt})$ and the expected value of Eq.~(\ref{kullexpecS1Sig}) is a measure of the goodness of the filtering procedure in keeping the maximal amount of information which can be present in sample correlation matrices estimated with a finite number of records.
   
 A second aspect concerns the stability of the filtered correlation matrix obtained from a sample matrix. Let us suppose to apply a certain filtering procedure to the correlation matrices ${\bf C}_1$ and ${\bf C}_2$ of two independent realizations of the system, obtaining two filtered correlation matrices ${\bf C}^{filt}_1$ and ${\bf C}^{filt}_2$. If it turns out that $K({\bf C}^{filt}_1,{\bf C}^{filt}_2)$ is larger than the expected value of  $K({\bf C}_1,{\bf C}_2)$ described by Eq.~(\ref{kullexpecS1S2}), one can conclude that the filtering procedure produces correlation matrices less reproducible than the sample correlation matrices and therefore the procedure is not suitable for the purpose of filtering robust information from the empirical correlation matrices ${\bf C}_1$ and ${\bf C}_2$.

In summary, in Tumminello et al. (2007a) we have shown that the Kullback-Leibler distance is very good for comparing correlation matrices. Its main properties are that (i) it is an asymmetric distance and therefore it can distinguish between quantities observed in real systems and used to model the empirical observations, e.g. the sample correlation matrix and the filtered correlation matrix respectively and 
(ii) the expectation values of the Kullback-Leibler distance given in Eq.s (\ref{kullexpecSigS1}-\ref{kullexpecS1S2}) are model independent, indicating that this distance is a good estimator of the statistical uncertainty due to the finite size of the empirical sample.
These properties are not observed in other widespread distances between matrices. For example, Tumminello et al. (2007a) have shown that these properties are not observed for the Frobenius distance, which is a standard measure of the distance between matrices. 

Biroli et al. (2007) have extended the above results on the Kullback-Leibler distance to a general class of elliptic distributions. 
Specifically, they considered random variables $x_i$ $(i=1,...,N)$ that can be generated by starting from random variables  $y_i$ $(i=1,...,N)$ following a generic multivariate distribution and by setting  $x_i=s y_i$, where $s$ is a positive random variable.

As a specific example, which is also relevant for financial data, they considered the multivariate Student's t-distribution. In this case the variables $y_i$ are taken from a multivariate normal distribution with correlation matrix ${\bf \Sigma}$ and $s$ is distributed according to 
\begin{equation}
P(s)=\frac{2}{\Gamma(\mu/2)}\exp\left[-\frac{s_0^2}{s^2}\right]\frac{s_0^\mu}{s^{1+\mu}}
\end{equation}
where $s_0^2=2\mu/(\mu-2)$ in such a way that $s$ has unit variance. The joint probability density function for the $x_i$ is
\begin{equation}
P(x_1,x_2,...,x_N)=\frac{\Gamma(\frac{N+\mu}{2})}{\Gamma(\mu/2)\sqrt{(\mu \pi)^N | {\bf \Sigma}|}}\frac{1}{\left(1+\frac{1}{\mu}\sum_{i,j} x_i( {\bf \Sigma}^{-1})_{ij}x_j\right)^{\frac{N+\mu}{2}}}
\label{studenteq}
\end{equation}
where the parameter $\mu$ gives the degrees of freedom of the distribution and describes the tail behavior of the marginal distribution of any $x_i$ since $P(x_i)\sim x_i^{-1-\mu}$. 


Let us assume that the correlation matrix is computed with the Pearson estimator for correlation coefficients.  Biroli et al. (2007) show that the Kullback-Leibler distance between two multivariate Student's t-distributions with the same scaling parameter $\mu$, and correlation matrices ${\bf \Sigma}_1$ and  ${\bf \Sigma}_2$ is
\begin{eqnarray}
\label{kullstud}
K({\bf \Sigma}_1,{\bf \Sigma}_2)=\frac{1}{2} \left[\log{\left(\frac{|{{\bf \Sigma}_2}|}{|{{\bf \Sigma}_1}|}\right)}+\nonumber\right.\\
\left.+(N+\mu)\int ds P(s) \log\left( \frac{1+\text{tr}\left({{\bf \Sigma}_2^{-1} {\bf \Sigma}_1 }\right)/(2 s)}{1+N/(2 s)}\right)\right],
\end{eqnarray}
In the limit $\mu/N \rightarrow \infty$ this expression coincides with the one obtained for the Gaussian case above, whereas in the limit $\mu/N \rightarrow 0$ Biroli et al. (2007) obtain
\begin{equation}
\label{kullstudsmallmu}
K({\bf \Sigma}_1,{\bf \Sigma}_2)=\frac{1}{2} \left[\log{\left(\frac{|{{\bf \Sigma}_2}|}{|{{\bf \Sigma}_1}|}\right)}+N\, \log\left( \frac{\text{tr}\left({{\bf \Sigma}_2^{-1} {\bf \Sigma}_1 }\right)}{N}\right)\right].
\end{equation}

Now consider a random vector $X$ of dimension $N$ with correlation matrix ${\bf \Sigma}$ and distributed according to Eq.~(\ref{studenteq}). Let ${\bf C}_1$ and ${\bf C}_2$ be two sample correlation matrices obtained from two independent realizations of the system both of length $T$.
It is possible to show (Biroli et al., 2007) that, similarly to the Gaussian case, the expectation values $E\left[K({\bf \Sigma},{\bf C}_1)\right]$, $E\left[K({\bf C}_1,{\bf \Sigma})\right]$ and $E\left[K({\bf C}_1,{\bf C}_2)\right]$, where the Kullback-Leibler distance is calculated according to either Eq.~(\ref{kullstud}) or Eq.~(\ref{kullstudsmallmu}), are not depending on ${\bf \Sigma}$, i.e. the expectation values are model independent.
This important property is valid not only for Gaussian multivariate variables but it holds true in general for elliptic multivariate distributions. This is due to the fact that for generic multivariate distributions the Kullback-Leibler distance can be written as $K({\bf \Sigma}_1,{\bf \Sigma}_2)=tr[f({\bf \Sigma}_2^-1 {\bf \Sigma}_1)]$, where $f$ is a function independent of the correlation structure of the system (Biroli et al. (2007)).

It is interesting to compare the expression of  Kullback-Leibler distance for Gaussian and Student's variables. Specifically, we can compare Eq. (\ref{kullbackGaussian}) and Eq. (\ref{kullstudsmallmu}). The right hand side of both the equations is the sum of two terms. The first term, $\frac{1}{2} \log{\left(\frac{|{{\bf \Sigma}_2}|}{|{{\bf \Sigma}_1}|}\right)}$, is the same for both the equations. Let's then focus our attention on the second terms of the equations and, in particular, on the second term of Eq. (\ref{kullstudsmallmu}), which is $\frac{1}{2} N\, \log\left[ \frac{1}{N}\text{tr}\left({{\bf \Sigma}_2^{-1} {\bf \Sigma}_1 }\right)\right]$. Suppose that the correlation matrix ${\bf \Sigma}_1$ is very close to ${\bf \Sigma}_2$, i.e. ${\bf \Sigma}_1\simeq{\bf \Sigma}_2$. Then, under this assumption,  $\frac{1}{N}\text{tr}\left({{\bf \Sigma}_2^{-1} {\bf \Sigma}_1 }\right)\simeq 1$. Under this assumption, at first order in $\frac{1}{N}\text{tr}\left({{\bf \Sigma}_2^{-1} {\bf \Sigma}_1 }\right)- 1$, we have
\begin{equation}
\label{approx}
\frac{N}{2} \, \log\left[ \frac{1}{N}\text{tr}\left({{\bf \Sigma}_2^{-1} {\bf \Sigma}_1 }\right)\right]\simeq \frac{N}{2}\left[ \frac{1}{N}\text{tr}\left({{\bf \Sigma}_2^{-1} {\bf \Sigma}_1 }\right)- 1\right],
\end{equation}
which is exactly the second term of the right hand side of Eq. (\ref{kullbackGaussian}). This calculation shows that whenever the correlation matrices involved in the Kullback-Leibler distance are very close one to the other (${\bf \Sigma}_1\simeq{\bf \Sigma}_2$) then the expression of the Kullback-Leibler distance for Student's t-distributions with small $\mu$ coincides, at the first order of approximation, with the expression of the Kullback-Leibler distance obtained for Gaussian random variables.

As a final remark we note that there is way to compute the expected value of a Kullback-Leibler distance which does not make use of the Pearson estimator. 
It is known that the Pearson's estimator of the correlation matrix is not the maximum likelihood estimator when the variables are non-Gaussian. In the case of the Student's t-distribution of Eq.~\ref{studenteq} there exists a recursive equation for the maximum likelihood estimator ${\bf \bar C}$ which is (Bouchaud and Potters, 2003)
\begin{equation}
\bar C_{ij}=\frac{N+\mu}{T}\sum_{t=1}^T\frac{x_i(t)x_j(t)}{\mu+\sum_{pq} x_p(t)({\bf \bar C}^{-1})_{pq}x_q(t)}
\label{mle}
\end{equation}
Since the maximum likelihood estimator of Eq. (\ref{mle}) of correlation matrix follows a Wishart distribution in the large $N$ limit, then the expectation values above can be calculated by making use of Wishart theory also in the case of Student's variable (Biroli et al., 2007; Tumminello et al., 2007b).

Finally, it is worth noting that there exists a straightforward modification of the HNFM of Eq. (\ref{model})  to generate hierarchically organized random variables with the multivariate Student's t-distribution. Specifically, we consider
\begin{equation}
\label{HNFMstud}
x_i (t)=\sum_{\alpha_h \in G(i)} {\gamma_{\alpha_h} f^{(\alpha_h)}(t)}\sqrt{\frac{\mu}{2 s(t)}}+\eta_i\,\, \epsilon_i (t)\sqrt{\frac{\mu}{2 s(t)}}.
\end{equation}

\section{Comparison of filtering procedures}
\label{sectionCOMP}

The KL distance can therefore be used to quantify and compare the performance of different filtering procedures of correlation matrices  (Tumminello et al., 2007a). A good filtering procedure should have two important properties: (i) being able to remove the ``right" amount of noise from the data in order to recover the signal and (ii) produce filtered matrices which are stable when one makes different observations of the same system. These two requirements are often in competition one with the other.  The proposed procedure to evaluate the performance of a filtering procedure is the following.

Suppose we are given with a data sample ${\bf X}$ and we have our favorite filtering procedure. We generate $M$ bootstrap replicas ${\bf X}_i$ ($i=1,..,M$) of the data. We then compute the sample correlation matrix ${\bf C}_i$ and apply the filtering procedure obtaining the filtered matrix ${\bf C}^{filt}_i$ to each replica ${\bf X}_i$. In order to measure the stability of the filtering procedure, we consider the average of the quantity $K({\bf C}^{filt}_i,{\bf C}^{filt}_j)$ over the replicas. An optimal filtering procedure should be perfectly stable (i.e. $\langle K({\bf C}^{filt}_i,{\bf C}^{filt}_j) \rangle=0$) because from each realization the filtering recovers the model matrix. 
In order to measure the filtered information we consider the average of $K({\bf C}_i,{\bf C}^{filt}_i)$ over the replicas. This quantity measures the information present in the sample correlation matrix ${\bf C}_i$ that has been discarded by the filtering procedure. 
We have seen above that for Gaussian variables the KL distance  $\langle K({\bf C}_i,{\bf \Sigma}) \rangle$ is different from zero and independent from the model ${\bf \Sigma}$ (see Eq.~(\ref{kullexpecS1Sig})). Therefore if our filtering procedure is recovering the true underlying model we should expect that $K({\bf C}_i,{\bf C}^{filt}_i)$ is equal to the right hand side of Eq. (\ref{kullexpecS1Sig}). We have thus a reference value for both the stability and the information expected from an optimal filtering and these values are independent from the underlying model. 
We represent the result of the analysis in a plane where the $x$ axis reports the stability $\langle K({\bf C}^{filt}_i,{\bf C}^{filt}_j) \rangle$ and the $y$ axis reports the information $\langle K({\bf C}_i,{\bf C}^{filt}_i)\rangle$.
In this plane the optimal point, labeled ${\bf \Sigma}$, has coordinate $x=0$ and $y$ equal to the right hand side of Eq. \ref{kullexpecS1Sig}. A filtering procedure will be considered good if the corresponding point in the stability-information plane is close to ${\bf \Sigma}$. 

To provide representative examples of quantitative analysis with the Kullback-Leibler distance of filtering procedures based on hierarchical clustering, as in the other sections, we consider SLCA and ALCA. We also consider filtering procedures based on the shrinkage technique (Ledoit and Wolf, 2003) and on the random matrix theory (RMT) (Laloux et al., 1999; Plerou et al., 1999; Rosenow et al., 2002; Potters et al., 2005). For the case of the the shrinkage procedure we construct a filtered matrix as
\begin{equation}
{\bf C}^{SHR}(\alpha)=\alpha {\bf T}+(1-\alpha){\bf C},
\label{shrinkage}
\end{equation}
where $0\le\alpha\le1$ and ${\bf T}$ is a target matrix. As commonly done in financial literature, we choose the target matrix as a matrix with $t_{ii}=1$ and $t_{ij}=\langle c_{ij}\rangle$ for $i\ne j$. We estimate the performance of the shrinkage procedure for different values of $\alpha$. It is interesting to note that there exist analytical methods to obtain the optimal value $\alpha^*$ according to a cost function  based on standard quadratic (or Frobenius) norm (Sch\"afer and Strimmer, 2005).  In the figures we also show the point (labeled ${\bf C}^{SHR}(\alpha^*)$) corresponding to the value $\alpha^*$.

In the econophysics literature, a widespread filtering procedure of the correlation matrix is based on the random matrix theory (Metha, 1990). If the $N$ variables are independent and with finite variance then in the limit $T,N \to \infty$, with a fixed ratio $Q\equiv T/N \geq 1$, the eigenvalues of the Pearson sample correlation matrix ${\bf C}$ is bounded from above by the value $ \lambda_{max}=\sigma^2 (1+1/Q+2\sqrt{1/Q})$ where $\sigma^2=1$ for correlation matrices. In some practical cases, such as for example in finance, one finds that the largest eigenvalue $\lambda_1$ of the empirical correlation matrix is definitely inconsistent with RMT. In these cases, Laloux et al. (1999) propose to modify the null hypothesis so that correlations can be explained in terms of a one factor model and $\sigma^2=1-\lambda_1/N$.
The  filtering procedure considered here has been proposed by Potters et al. (2005) and it works as follows. One diagonalizes the correlation matrix and replaces the all eigenvalues smaller than   $\lambda_{max}$ in the diagonal matrix with their average value. Then one retransforms the modified diagonal matrix in the standard basis obtaining a matrix ${\bf H}_{RMT}$ of elements $h_{ij}^{RMT}$ to preserve the trace. Finally, the filtered correlation matrix ${\bf C}^{RMT}$ is the matrix of elements $c_{ij}^{RMT}=h_{ij}^{RMT}/\sqrt{h_{ii}^{RMT}\,h_{jj}^{RMT}}$.

\begin{center}
\begin{figure}[ptb]

\begin{center}
              \hbox{
              \includegraphics[scale=0.25]{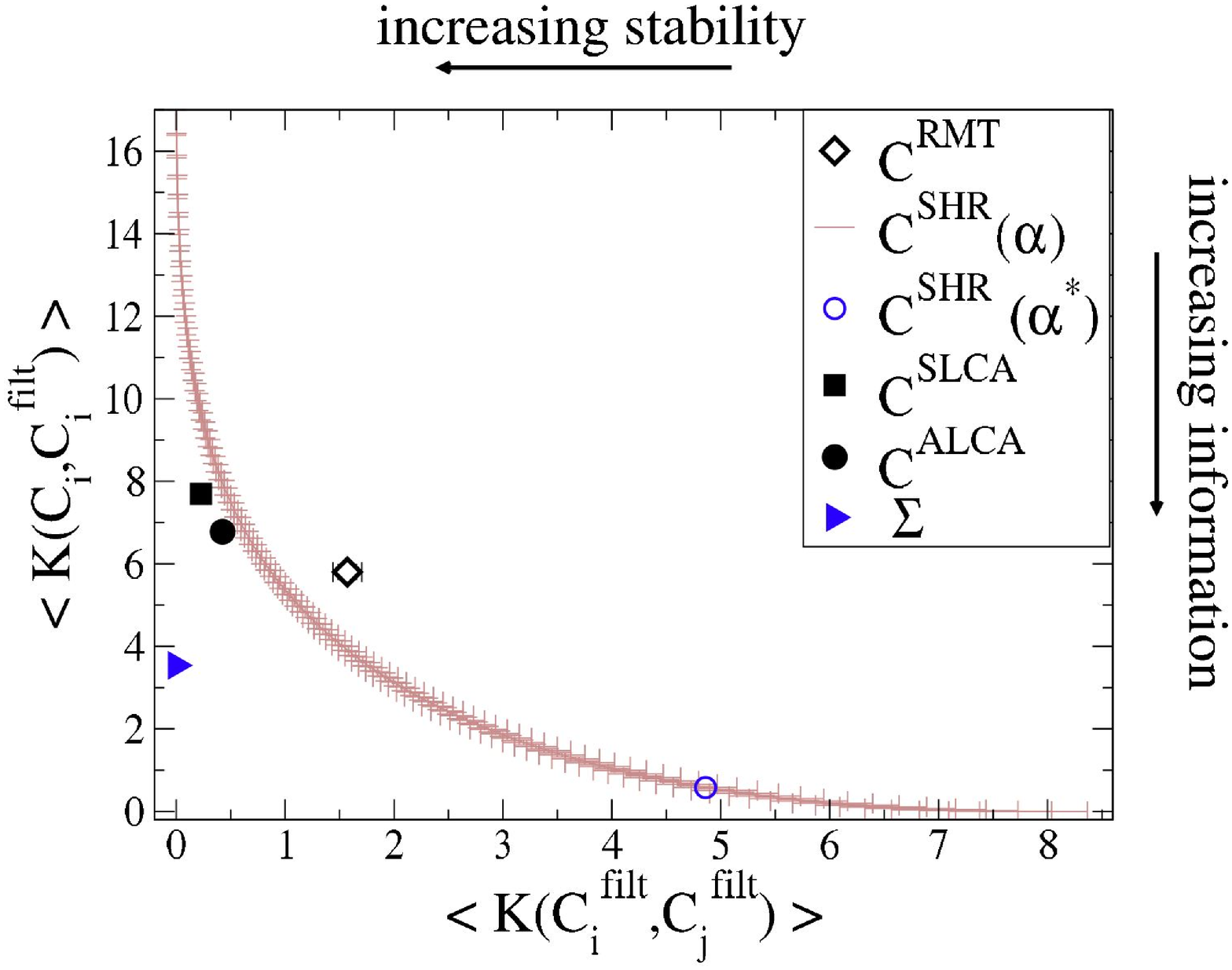}
              \hspace{. cm}
              \includegraphics[scale=0.25]{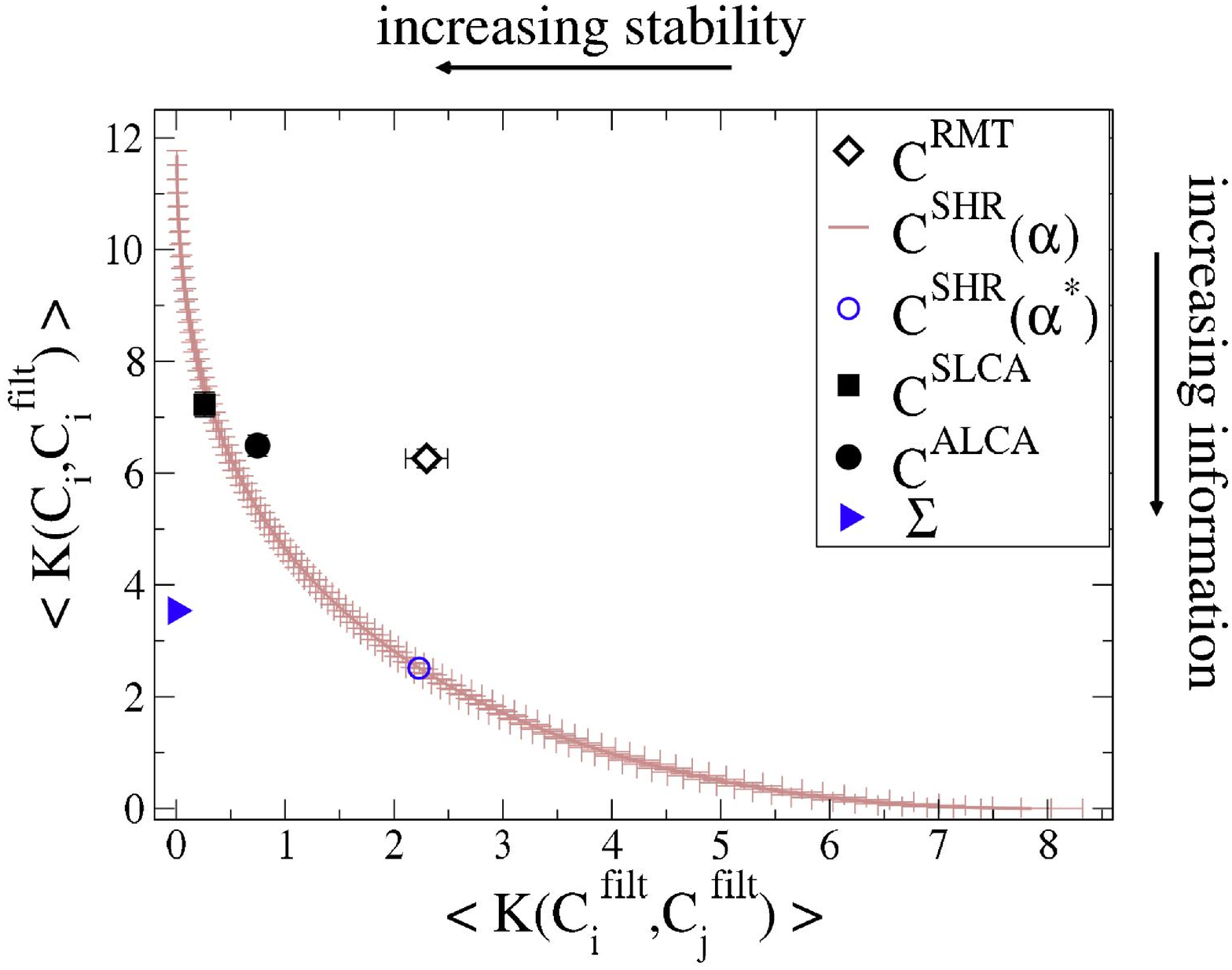}
              }
\end{center}             
 \caption{Stability of the filtered matrix ($x$ axis) against the amount of information about the correlation matrix that is retained in the filtered matrix ($y$ axis).  The points labeled with ${\bf C}^{SHR}(\alpha)$ correspond to the shrinkage procedure (see Eq.~\ref{shrinkage}) and the parameter $\alpha$ goes from $0$ to $1$ when one goes from the bottom right to the top left corner. Left panel shows the result for a block diagonal model of $N=100$ elements divided in $12$ groups and simulated for $T=748$ points. Right panel shows the result for a hierarchically nested model of 100 elements following the HNFM with 23 factors of Tumminello et al. (2007d).} 
 \label{RMT}
\end{figure}
\end{center}

In fig.~\ref{RMT} we show the KL distance in the plane stability-information for these filtering procedures applied to artificial data generated according to two different models. The left panel shows the result for a model whose correlation matrix is block diagonal with $12$ blocks, whereas the right panel shows the result for a HNFM with $23$ factors. In both panels we show the points corresponding to the RMT, SLCA, and ALCA filtering procedures. 
We also show the points corresponding to the shrinkage filtering procedure of Eq.~\ref{shrinkage} for different values of $\alpha$.
The shrinkage method is capable to achieve a very good compromise between stability and information. From this analysis it is possible to extract an optimal value of $\alpha$ minimizing the distance from the point labeled with $\Sigma$. It should be noted that this value in general does not coincide with the value $\alpha^*$ obtained with the method of Sch\"afer and Strimmer (2005), i.e. by minimizing the Frobenius norm.

\begin{figure}[ptb]
\begin{center}
              \includegraphics[scale=0.5]{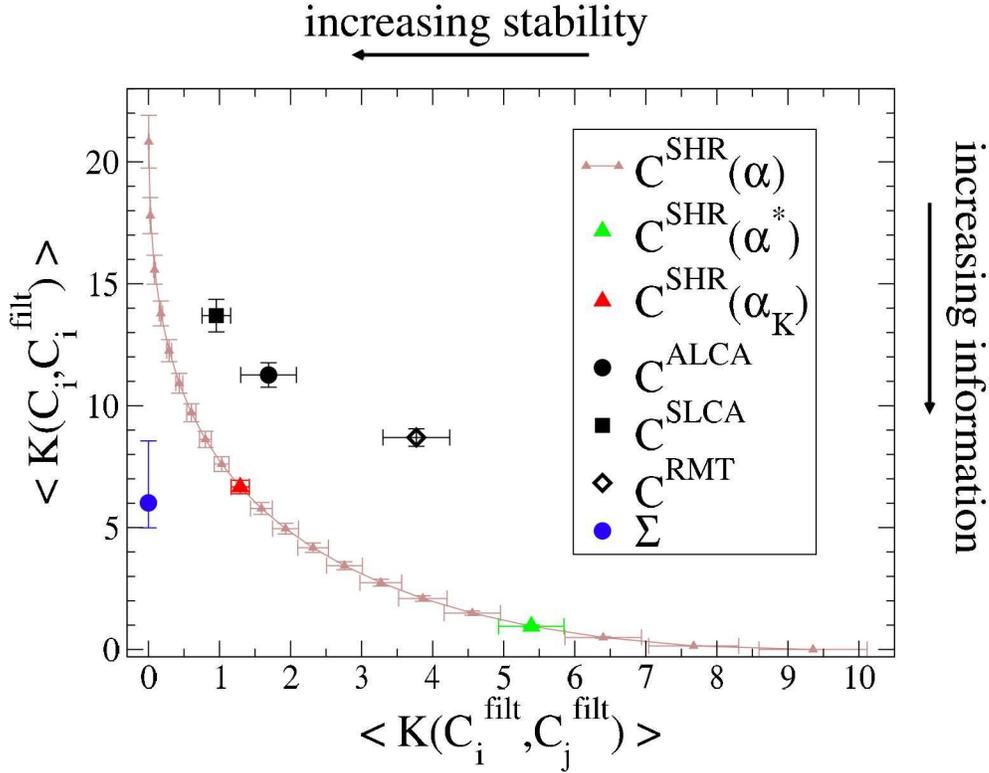}
              \caption{Stability of the filtered matrix ($x$ axis) against the amount of information about the correlation matrix that is retained in the filtered matrix ($y$ axis) for $N=100$ stocks of the NYSE in the period 2001-2003 ($T=748$). The points labeled with ${\bf C}^{SHR}(\alpha)$ correspond to the shrinkage procedure (see Eq.~\ref{shrinkage}) and the parameter $\alpha$ goes from $0$ to $1$ when one goes from the bottom right to the top left corner. The point ${\bf C}^{SHR}(\alpha_K)$ is obtained for $\alpha=\alpha_K=0.55$, which is the value of the shrinkage parameter that minimize the euclidean distance in the plane stability-information between the curve corresponding to ${\bf C}^{SHR}(\alpha)$ and the point associated with $\Sigma$. The latter point represents the expectation value for a filtering procedure able to recover the true correlation matrix of the system. See the text for more details.} 
              \label{NYSE}
\end{center}
\end{figure}

We now consider an application to a real system. We investigate the daily returns of $N=100$ highly capitalized stocks traded at the NYSE in the period 2001-2003 ($T=748$). In present study, differently than in Tumminello et al. (2007a) where we worked in the Gaussian approximation, we assume stock returns to be Student's t-distributed according to Eq. (\ref{studenteq}), . The scaling parameter $\mu$ in the distribution of Eq. (\ref{studenteq}) is assumed to be the same for all of the stocks in the portfolio. Accordingly, we determine $\mu$ as the average value of the maximum likelihood estimates of $\mu$ independently evaluated for each stock. The result is $\mu=5.9$ with a standard deviation $\sigma_{\mu}=1.8$. The ratio $\mu/N=0.059$ is much less than one and therefore ensures that Eq. (\ref{kullstudsmallmu}) is a good approximation of the Kullback-Leibler distance for the present case. In Fig. ~\ref{NYSE} we show  the performance of different filtering procedures in the plane stability-information. In the figure the Kullback-Leibler distance is calculated according to Eq. (\ref{kullstudsmallmu}).

The filtering procedures based on RMT, SLCA and ALCA have different properties in terms of stability and information (Tumminello et al., 2007a). SLCA is the most stable even if it is the least informative, whereas RMT is the least stable but the most informative. ALCA has intermediate properties both with respect to stability and to information.
The filtering procedure based on shrinkage seems to outperform the other filtering techniques for selected values of the $\alpha$ parameter. In Fig.~\ref{NYSE}, we also highlight (red color)  a point corresponding to the optimal value $\alpha_K=0.55$ of the shrinkage parameter $\alpha$ in the plane stability-information. 
This point is obtained by minimizing the Euclidean distance between the points $(0,\,<K({\bf C}_i,{\bf\Sigma})>)$ 
and $(<K({\bf C}^{SHR}_i(\alpha),{\bf C}^{SHR}_j(\alpha))>,\,<K({\bf C}_i,{\bf C}^{SHR}_j(\alpha))>)$ in the space stability-information. The point $(0,\,<K({\bf C}_i,{\bf\Sigma})>)$ is expected for a filtering procedure able to perfectly detect the correlation matrix ${\bf \Sigma}$ of the system. Our aim is now to estimate $<K({\bf C}_i,{\bf\Sigma})>$. By using the estimate of $\mu$, we evaluate the quantity $<K({\bf C}_i,{\bf\Sigma})>$ by exploiting the model independency of the Kullback-Leibler distance. During our analysis we realized that for Student's t-distributions the bootstrap replicas, which we actually use to deal with real data, can introduce a bias with respect to independent simulations. By performing numerical simulations of Student's t-distributed random variables characterized by different values of $\mu$, we notice that this bias does not appear in the case of normal or close to normal distributions (typically when $\mu$ is greater than 10). In order to overcame the problem when Student's t-distributed random variables with a low value of $\mu$ describe the data better than Gaussian random variables, we perform 100 independent simulations of Student's t-distributed data series of length $T=748$ (the same as real data) according to the model  of Eq.  (\ref{HNFMstud}) with scaling parameter $\mu=5.9$ and we construct 100 bootstrap replicas of each simulated data series. It is to notice that the choice of  Eq.  (\ref{HNFMstud}) does not affect the generality of results because the expectation values of the Kullback-Leibler distance do not depend on the correlation structure of the model.
We indicate the correlation matrix of simulated series with ${\bf C}_j$ ($j=1,...,100$), and the correlation matrix of a bootstrap replica associated with ${\bf C}_j$ with ${\bf C}_{ji}^b$ (i=1,...,100).
Because the expectation value of the correlation matrix ${\bf C}_{ji}^b$ is ${\bf C}_j$ and the expectation values of the Kullback-Leibler distance are model independent, we can estimate the value of $<K({\bf C}_i,{\bf\Sigma})>$ as $<K({\bf C}_{ji}^b,{\bf C}_i)>$, where the average is taken over both the indices $i$ and $j$.  The result we obtain $<K({\bf C}_{ji}^b,{\bf C}_i)>=6.01$. This result is shown in the Fig.~\ref{NYSE} as a blue circle. Finally, in order to associate an error bar with this value, we apply the same procedure used to obtain the estimate of $<K({\bf C}_i,{\bf\Sigma})>$ for values of $\mu$ equal to $\mu_{min}=\mu-\sigma_{\mu}=5.9-1.8=4.1$ (providing the value at the top of the error bar in the figure) and $\mu_{max}=\mu+\sigma_{\mu}=5.9+1.8=7.7$ (providing the value at the bottom of the error bar in the figure). A series of numerical simulations performed with different model generating the dynamics of the random variables show that the bias introduced by bootstrapping data instead of performing independent simulations turns out to be approximately independent of the actual correlation matrix of the system and for $\mu=5.9$, N=100, and $T=748$ the bias is equal to $-10.8\%\pm1.7\%$, i.e. the bootstrap based estimation gives an underestimate of $<K({\bf C}_i,{\bf\Sigma})>$. However, in the investigations summarized in  Fig.~\ref{NYSE} the bias is the same for all of the points and therefore the comparison of filtering procedure is possible. Our numerical simulations also show that the value of the bias tends to increase as the value of $\mu$ decreases. 

It is also worth noting that results very similar to those shown in Fig.~\ref{NYSE} can be obtained by using the expression of Eq.  (\ref{kullbackGaussian}) (valid for Gaussian variables) to evaluate the Kullback -Leibler distance instead of Eq. (\ref{kullstudsmallmu}) (valid for Student's t-distributed variables with  $\mu/N <<1$). This fact can be interpreted by observing that the correlation matrices involved in the Kullback-Leibler distance do not differ much one from the other and, therefore, Eq. (\ref{kullstudsmallmu}) gives estimates of the Kullback-Leibler distance similar to those obtained by using Eq. (\ref{kullbackGaussian}) that is strictly speaking only valid for Gaussian variables, and which has been used in Tumminello et al. (2007a). 
A final remark concern the shrinkage technique. We note that the shrinkage parameter $\alpha_K=0.55$ is significantly different than $\alpha^*=0.16$, obtained by minimizing the Frobenius norm. The point associated with $\alpha^*=0.16$ (green circle in Fig.~\ref{NYSE}) in the plane stability-information is quite far from the point of an ideal filtering procedure able to perfectly detect the correlation matrix ${\bf \Sigma}$ of the system (blue circle in Fig.~\ref{NYSE}). This observation suggests that by using the Frobenius distance to get an estimate of the shrinkage parameter $\alpha$ one puts too much faith on the statistical robustness of the sample correlation matrix. Conversely, $\alpha_K$ represents a more reliable estimate of the optimal shrinkage parameter.

\section{Conclusions}
\label{sectionCONC}

This paper discusses several methods to quantitatively investigate the properties of the correlation matrix of a system of N elements. In the present work we consistently investigate the correlation matrix of the synchronous dynamics of the returns of a portfolio of financial assets. However, our results apply to any correlation matrix computed starting from the series of T records belonging to N elements of a system of interest. 

Specifically, we discuss how to associate to a correlation matrix a hierarchical tree and correlation based trees or graphs. In previous papers we have shown that the information selected through these clustering procedures and the construction of correlation based trees or graphs are pointing out interesting details on the investigated system. For example, the hierarchical clustering is able to detect clusters of stocks belonging to the same sectors or sub-sectors of activities without the need of any supervision of the clustering procedure. We have also shown that the information present in correlation based trees and graphs provides additional clues about the interrelations among stocks of different economic sectors and sub-sectors. It is worth noting that this kind of information is not present in the information stored in the hierarchical trees obtained by ALCA and SLCA clustering procedures or, equivalently, in the associated ultrametric correlation matrices. The information obtained from what we call the ``filtering procedure" of the correlation matrix is subjected to statistical uncertainty. For this reason, we discuss a bootstrap methodology able to quantify the statistical robustness of both the hierarchical trees and correlation based trees or graphs.   

The hierarchical trees and correlation based trees and graphs associated with portfolios of stocks traded in financial markets often show clusters of stocks partitioned in sub-clusters, sub-clusters partitioned in sub-sub-clusters and so on until the level of the single stock. The ubiquity of this observation has motivated us to develop a hierarchically nested factor model able to fully describe this property. Our model is a nested factor model characterized by the same correlation matrix as the empirical set of data. The model is expressed in a direct and simple form when all the correlation coefficients are positive or very close to positive (for a precise definition of the limits of validity of this extension see Section \ref{sectionH}). The number of factors of the model is by construction equal to the number of elements of the system. Again the selection of the most statistically reliable factors detected in a real system is obtained by a procedure based on bootstrap with a bootstrap threshold selected in a self-consistent way. 

The amount of information and the statistical stability of filtering procedures of the correlation matrix are quantified by using the Kullback-Leibler distance. We report and discuss analytical results both for Gaussian and for Student's t-distributed multivariate time series. In both cases the expectation values of the Kullback-Leibler distance are model independent, indicating that this distance is a good estimator of the statistical uncertainty due to the finite size of the empirical sample. These properties are not observed in other widespread distances between matrices such as, for example , the Frobenius distance, which is a standard measure of the distance between matrices. In our example with real data, we estimate the amount of information retained and the stability of the filtering procedure used in a data set of 100 stocks approximately described by a multivariate Student's $t$ distribution. For this set of data we are able to discriminate among filtering procedures as different as ALCA, SLCA, random matrix theory and a shrinkage procedure.


{\bf Acknowledgments} We acknowledge partial support from PRIN-MIUR research project ``Progettazione di mercati e modelli ad agenti di comportamento di trading''.






\newpage

\begin{table}\nonumber
\caption{1A- Stocks with tick symbol from ABT to IGT. The first column is the tick symbol in alphabetical order, the second column reports an abbreviation of the economic sector of the considered company. Specifically, we have Basic Materials (BM), Consumer Cyclical (CC), Consumer Non Cyclical (CNC), Energy (E), Services (S), Financial (F), Healthcare (H), Technology (T), Utilities (U), Transportation (TR), Conglomerates (CO) and Capital Goods (CG). The third column indicates the economic sub-sector of the company. The forth column reports the company name whereas the fifth column is the numerical label of the stock used in Fig.s \ref{dendroData} and \ref{nodered}.}
\begin{tabular}{||l|c|c|c|c||}
\hline
{\bf tick} & {\bf sector} & {\bf sub-sector} & {\bf name} & {\bf ord}\\
\hline
\hline
ABT & H & Major Drugs & Abbott Laboratories            & 72 \\
ADI & T & Semiconductors & Analog Devices Inc             & 48 \\
AFL & F & Insurance Accidental \& Health & Aflac Inc                      & 62 \\
AIG & F & Insurance Prop. \& Casualty & American Intl Group Inc        & 16 \\
ALL & F & Insurance Prop. \& Casualty & Allstate Corp  The             & 60 \\
AVP & CNC & Personal \& Household Products & Avon Products Inc              & 82 \\
AXP & F & Consumer Financial Services & American Express Company       & 6 \\
BA & CG & Aerospace \& Defense & Boeing Co                      & 36 \\
BAC & F & Money Center Banks & Bank Of America Corp           & 12 \\
BAX & H & Medical Equipment \& Supplies & Baxter International Inc       & 78 \\
BBY & S & Retail Technology & Best Buy Co Inc                & 43 \\
BK & F & Money Center Banks & Bank Of New York Inc           & 14 \\
BLS & S & Communication Services & Bellsouth Corporation          & 67 \\
BMY & H & Major Drugs & Bristol Myers Squibb Company   & 74 \\
BNI & TR & Railroad & Burlington Nrthrn Santa Fe Com & 38 \\
BSC & F & Investment Services & Bear Stearns Companies Inc     & 2 \\
BSX & H & Medical Equipment \& Supplies & Boston Scientific Corp         & 97 \\
BUD & CNC & Beverages Alcoholic & Anheuser Busch Cos Inc         & 87 \\
CA & T & Software \& Programming & Computer Associates Intl Inc   & 52 \\
CAG & CNC & Food Processing & Conagra Foods Inc.             & 91 \\
CAH & H & Biotechnology \& Drugs & Cardinal Health Inc            & 75 \\
CAT & CG & Constr. \& Agric. Machinery & Caterpillar Inc                & 29 \\
CCU & S & Broadcasting \& Cable TV & Clear Channel Communictns Inc  & 22 \\
CI & F & Insurance Accidental \& Health & Cigna Corp                     & 63 \\
CL & CNC & Personal \& Household Products & Colgate-palmolive Co           & 80 \\
DD & BM & Chemical - Plastic \& Rubber & Du Pont De Nemours E I  Co     & 25 \\
DE & CG & Constr. \& Agric. Machinery & Deere   Co                     & 30 \\
DHR & T & Scientific \& Technical Instr. & Danaher Corp                   & 32 \\
DIS & S & Broadcasting \& Cable TV & Walt Disney Co-disney Common   & 21 \\
DOW & BM & Chemical - Plastic \& Rubber & Dow Chemical Co                & 27 \\
DUK & U & Electric Utilities & Duke Energy Corporation        & 93 \\
\hline
\hline
\end{tabular}
\end{table}
\label{table1A}

\begin{table}\nonumber
\caption{1B- Stocks with tick symbol from EMC to LOW. The content of columns is the same as in Table 1A.}
\begin{tabular}{||l|c|c|c|c||}
\hline
{\bf tick} & {\bf sector} & {\bf sub-sector} & {\bf name} & {\bf ord}\\
\hline
\hline
EMC & T & Computer Storage Devices & Emc Corporation                & 49 \\
EMR & CO & Conglomerates & Emerson Electric Co            & 33 \\
FDC & T & Computer Services & First Data Corp                & 46 \\
FNM & F & Consumer Financial Services & Fannie Mae                     & 58 \\
FON & S & Communication Services & Sprint Corp Fon Group          & 68 \\
FRE & F & Consumer Financial Services & Freddie Mac D/b/a Voting       & 59 \\
G & CNC & Personal \& Household Products & Gillette Co                    & 83 \\
GCI & S & Printing \& Publishing & Gannett Co Inc                 & 19 \\
GD & CG & Aerospace \& Defense & General Dynamics Corp          & 95 \\
GDT & H & Medical Equipment \& Supplies & Guidant Corp                   & 98 \\
GDW & F & S\&Ls/Savings Banks & Golden West Financial Corp     & 61 \\
GE & CO & Conglomerates & General Electric & 5 \\
GIS & CNC & Food Processing & General Mills Inc              & 90 \\
GM & CC & Auto \& Truck Manufacturers & General Motors Corp            & 34 \\
GPS & S & Retail Apparel & Gap Inc  The                   & 45 \\
HD & S & Retail Home Improvement & Home Depot Inc                 & 39 \\
HDI & CC & Recreational Products & Harley Davidson Inc            & 24 \\
IBM & T & Computer Hardware & Intl Business Machines Corp    & 50 \\
IGT & S & Casinos \& Gaming & Intl Game Technology           & 56 \\
IP & BM & Paper \& Paper Products & International Paper Co         & 28 \\
ITW & CG & Misc. Capital Goods & Illinois Tool Works            & 31 \\
JNJ & H & Major Drugs & Johnson And Johnson            & 71 \\
K & CNC & Food Processing & Kellogg Co                     & 89 \\
KMB & BM & Paper \& Paper Products & Kimberly Clark Corp            & 81 \\
KO & CNC & Beverages Non-Alcoholic & Coca-cola Co                   & 84 \\
KR & S & Retail Grocery & Kroger Co                      & 94 \\
KRB & F & Regional Banks & M B N A Corp                   & 7 \\
KSS & S & Retail Department \& Discount & Kohls Corp                     & 42 \\
LEH & F & Investment Services & Lehman Brothers Holdings       & 1 \\
LLY & H & Major Drugs & Lilly Eli   Co                 & 73 \\
LOW & S & Retail Home Improvement & Lowes Companies Inc            & 40 \\
\hline
\hline
\end{tabular}
\end{table}
\label{table1B}

\begin{table}\nonumber
\caption{1C- Stocks with tick symbol from MCD to WMT. The content of columns is the same as in Table 1A.}
\begin{tabular}{||l|c|c|c|c||}
\hline
{\bf tick} & {\bf sector} & {\bf sub-sector} & {\bf name} & {\bf ord}\\
\hline
\hline
MCD & S & Restaurants & Mcdonalds Corp                 & 99 \\
MDT & H & Medical Equipment \& Supplies & Medtronic Inc                  & 77 \\
MEL & F & Investment Services & Mellon Financial Corp          & 15 \\
MER & F & Investment Services & Merrill  Lynch   Co Inc        & 3 \\
MMC & F & Insurance Miscellaneous & Marsh   Mclennan Cos Inc       & 18 \\
MOT & T & Communication Equipment & Motorola Inc                   & 51 \\
MRK & H & Major Drugs & Merck   Co Inc                 & 70 \\
NCC & F & Regional Banks & National City Corp             & 8 \\
NEM & BM & Gold \& Silver & Newmont Mining Corp  Holding C & 100 \\
NOC & CG & Aerospace \& Defense & Northrop Grumman Cp  Hldg Co   & 96 \\
OMC & S & Advertising & Omnicom Group Inc              & 23 \\
ONE & F & Regional Banks & Bank One Corp                  & 10 \\
OXY & E & Oil \& Gas Operations & Occidental Petroleum Corp      & 54 \\
PEP & CNC & Beverages Non-Alcoholic & Pepsico Inc                    & 85 \\
PFE & H & Major Drugs & Pfizer Inc  & 69 \\
PG & CNC & Personal \& Household Products & Procter   Gamble Co            & 79 \\
PGR & F & Insurance Prop. \& Casualty & Progressive Corp               & 17 \\
PNC & F & Regional Banks & Pnc Finl Svcs Grp Inc The      & 11 \\
PPG & BM & Chemical Manifacturing & Ppg Industries Inc             & 26 \\
RD & E & Oil \& Gas - Integrated & Royal Dutch Pet  New 1.25gldrs & 55 \\
S & S & Retail Department \& Discount & Sears Roebuck Co               & 44 \\
SBC & S & Communication Services & Sbc Communications Inc         & 66 \\
SCH & F & Investment Services & Schwab Charles Corp            & 4 \\
SGP & H & Major Drugs & Schering Plough Corp           & 76 \\
SLB & E & Oil Well Services \& Equipment & Schlumberger Ltd               & 53 \\
SLE & CNC & Food Processing & Sara Lee Corp                  & 88 \\
SO & U & Electric Utilities & Southern Co                    & 92 \\
STI & F & Regional Banks & Suntrust Banks Inc             & 9 \\
SYY & S & Retail Grocery & Sysco Corp                     & 86 \\
TRB & S & Printing \& Publishing & Tribune Company                & 20 \\
TXN & T & Semiconductors & Texas Instruments              & 47 \\
TYC & CO & Conglomerates & Tyco International Ltd  New    & 65 \\
UNP & TR & Railroad & Union Pacific Corporation      & 37 \\
UTX & CO & Conglomerates & United Technologies Corp       & 35 \\
WAG & S & Retail Drugs & Walgreen Company               & 57 \\
WFC & F & Money Center Banks & Wells Fargo   Co New           & 13 \\
WLP & F & Insurance Accidental \& Health & Wellpoint Hlth Netwks Hldg Co  & 64 \\
WMT & S & Retail Department \& Discount & Wal-mart Stores Inc    & 41 \\
\hline
\hline
\end{tabular}
\end{table}
\label{table1C}

\newpage

\end{document}